\documentclass[useAMS,usenatbib]{mn2e}
\usepackage{epsfig}
\title{Kinematics, Substructure and Luminosity-weighted Dynamics of Six Nearby Galaxy Groups}
\author[Firth et al.]
       {\parbox{16cm}{P. Firth\thanks{E-mail: firth@physics.uq.edu.au}$^{1}$, E. Evstigneeva$^{1}$, J. B. Jones$^{2}$, M. J. Drinkwater$^{1}$,\\
       S. Phillipps$^{3}$ and M. D. Gregg$^{4}$}\\
        \\
       ${1}$ Department of Physics, University of Queensland, Brisbane, Qld 4072, Australia.\\
       ${2}$ Astronomy Unit, School of Mathematical Sciences,  Queen Mary, University of London, Mile End Road, London E1 4NS, U.K.\\
       ${3}$ Astrophysics Group, H.H. Wills Physics Laboratory, University of Bristol, Tyndall Avenue, Bristol BS8 1TL, U.K.\\
       ${4}$ Department of Physics, University of California, One Shields Avenue, Davis, CA 95616-8677, U.S.A. }
\date{Accepted 2006 April 1, Received 2006 March 1}
\pagerange{\pageref{firstpage}--\pageref{lastpage}} 
\pubyear{2006}

\begin{document}

\bibliographystyle{mn2e}

\maketitle

\label{firstpage}

\begin{abstract}
We have redefined group membership of six southern galaxy groups in the local universe (mean cz $< 2000 \; \mbox{km} \, \mbox{s}^{-1}$) based on new redshift measurements from our recently acquired Anglo--Australian Telescope 2dF spectra. For each group, we investigate member galaxy kinematics, substructure, luminosity functions and luminosity-weighted dynamics. Our calculations confirm that the group sizes, virial masses and luminosities cover the range expected for galaxy groups, except that the luminosity of NGC 4038 is boosted by the central starburst merger pair. We find that a combination of kinematical, substructural and dynamical techniques can reliably distinguish loose, unvirialised groups from compact, dynamically relaxed groups.  Applying these techniques, we find that Dorado, NGC 4038 and NGC 4697 are unvirialised, whereas NGC 681, NGC 1400 and NGC 5084 are dynamically relaxed. 
\end{abstract}


\section{Introduction}

Galaxy groups are important structures, containing a significant fraction of the galaxy population in the universe. In November 2004 (ATAC/2004B/19) and April 2005 (PATT/2005A/13), we observed six nearby southern galaxy groups (Dorado, NGC 681, NGC 1400, NGC 4038, NGC 4697 and NGC 5084) with the 2dF multi-object spectrograph at the 3.9-m Anglo--Australian Telescope. The primary purpose of these observations was to locate ultra-compact dwarf galaxies (UCDs). In this paper we combine non-UCD galaxy redshift measurements from these observations with previously catalogued data to create revised group membership lists. We investigate luminosity-weighted group kinematics and dynamics based upon the virial method employed in \citet{Ferguson..1990}, using Monte Carlo modelling to estimate uncertainties due to observational errors.

A meaningful kinematical and dynamical analysis of galaxy groups requires a clear definition of group membership -- but galaxy groups are variously defined depending on the group detection algorithm selected. Researchers have historically relied on positional, morphological and surface brightness information to define galaxy groups, but increasing access to redshift data has improved group membership definition. For example, \citet{Huchra..1982} applied a ``friends of friends'' percolation algorithm to search for iso-density contours in redshift space, and \citet{Maia..1989} used the Southern Sky Redshift Survey to produce a Catalog of Southern Groups of Galaxies. \citet{Ferguson..1990} used galaxy morphology, together with available but limited redshifts, to create catalogs of nearby galaxy groups, including Dorado and NGC 1400. \citet{Garcia..1993} combined the percolation method of \citet{Huchra..1982} with a hierarchical method derived from \citet{Tully..1987} to create a large catalogue of galaxy groups, including NGC 681, NGC 4038, NGC 4697 and NGC5084. Starting with group membership as defined by \citet{Ferguson..1990} and \citet{Garcia..1993}, our results more accurately define these six galaxy groups as gravitationally-bound systems in redshift space.

The structure of our paper is as follows -- Section (2) describes our 2dF observations and redshift results obtained; Section (3) details the methods, results and interpretation we have used in our analysis of group kinematics, substructure and luminosity-weighted dynamics; and Section (4) summarises our findings.

\section{Observations}

\subsection{Description of Galaxy Groups Observed}

The groups selected for our 2dF observations are low redshift (mean cz $\leq 2000 \; \mbox{km} \, \mbox{s}^{-1}$), southern sky galaxy groups, with a central dominant galaxy and a broad density range:

 \begin{itemize}
 \item The Dorado Group has been variously identified by past researchers as Shk18 \citep{Shakhbazian..1957}, G16 \citep{deVaucouleurs..1975}, HG3 \citep{Huchra..1982} and MC13 \citep{Maia..1989}. De Vaucouleurs described it as `a large and complex nebula in the region of Dorado...part of a larger cloud complex'. It is described by \citet{Ferguson..1990} as `a loose concentration of spirals and ellipticals', and later by \citet{Carrasco..2001} as covering an area of 10 degrees square. The dominant galaxies are, in order of decreasing luminosity, NGC 1566 (spiral), NGC 1553 (lenticular) and NGC 1549 (elliptical).

 \item The NGC 681 Group is a sparse group of galaxies, stretching across the 2dF field and centred about half-way between the two brightest barred-spiral galaxies -- the starburst galaxy NGC 701 and the slightly fainter galaxy NGC 681.
 
 \item The NGC 1400 Group is `a loose group of galaxies that is part of the larger Eridanus cloud complex' \citep{Ferguson..1990}. Ferguson and Sandage identified 120 galaxies as potential members of this group. The brightest member is the E0 elliptical NGC 1407, followed by the spiral NGC 1400. These two galaxies have a considerable radial velocity difference, although they are considered to be at a similar distance, which is interpreted as evidence of a large dark matter halo \citep{Quintana..1994}.
 
 \item There are 27 galaxies in the NGC 4038 Group as listed by \citet{Garcia..1993}, including the merging pair of spiral galaxies NGC 4038 and NGC 4039, collectively known as the Antennae Galaxy and featuring in the pioneering N-body simulations of \citet{Toomre..1972}.
 
 \item The NGC 4697 Group contains 19 members identified by \citet{Garcia..1993} and dominated by NGC 4697(elliptical). 
 
 \item The NGC 5084 Group comprises only 5 galaxies identified by \citet{Garcia..1993} and is dominated by the lenticular galaxy NGC 5084.
 
 \end{itemize}

\subsection{Target Selection}

The primary aim of our 2dF spectroscopic survey was to discover whether UCDs exist in galaxy groups, and the results of this research will be published separately. We therefore selected point-source targets in each group as our primary targets for spectroscopy. However, candidate galaxy members of these groups were also targeted, by allocating a small proportion of the available 2dF fibres, to refine information about the group structures and dynamics to aid in interpreting UCD results. The results from these galaxy redshift observations are presented in this paper.

\begin{figure}
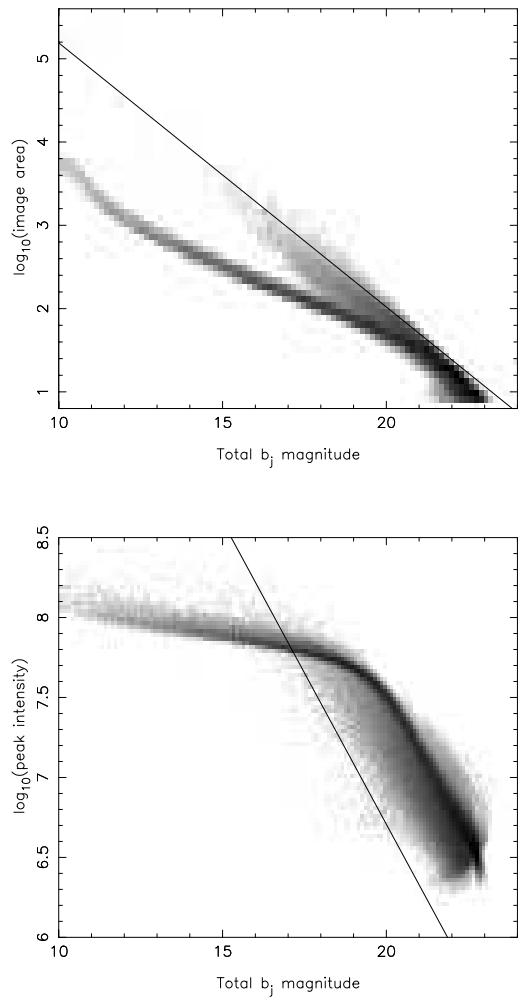
   \centering
% fig1a.eps = fig_btot_logarea_ngc4038.eps
% fig1b.eps = fig_btot_logipeak_ngc4038.eps
 \includegraphics[width=7cm, angle=-90]{fig1a.eps}
 \includegraphics[width=7cm, angle=-90]{fig1b.eps}
 \caption{Morphological selection of likely group member galaxies -- typical plots for galaxy group NGC 4038.  The data was taken from the SuperCOSMOS Sky Survey catalogue for the blue band. Objects classified as galaxies form the lighter gray background to the Galactic stars in the narrow, dark grey band.  UPPER: Total magnitude plotted against image area. Previous results of UCD searches in the Fornax galaxy cluster indicate that group member galaxies lie above the line shown.  LOWER: Total magnitude plotted against peak intensity. Previous results show that group member galaxies lie below the line shown, excluding the stellar locus in the upper part which cuts across due to saturation.}
 \label{fig:UCDTargets}
 \end{figure}

For the NGC 681, NGC 4038, NGC 4697 and NGC 5084 groups, catalogued by \citet{Garcia..1993}, we selected candidate group galaxies using size and surface brightness characteristics of the galaxy images. Using blue band data from the SuperCOSMOS Sky Survey catalogue, the positions of all objects in the 2dF fields were plotted in diagrams of isophotal area (at the detection threshold surface brightness) against total magnitude, and of the SuperCOSMOS peak intensity against total magnitude (see Fig. \ref{fig:UCDTargets} for examples). Similar data in the region of the Fornax Cluster were used to define the regions in these planes in which cluster galaxies are located, using the member galaxy list of \citet{Ferguson..1990}. Extending these Fornax Cluster results to the galaxy groups, regions were defined where there was likely to be a high probability of finding group galaxies but a modest probability of finding background galaxies. Images of the objects found in these regions were inspected visually to reject any obvious haloes and diffraction spikes of stars, plate flaws, and sub-regions of bright galaxies. This left samples of candidate group galaxies for targeting with 2dF fibres.

For the NGC 1400 and Dorado galaxy groups, targets were selected from the \citet{Ferguson..1990} catalogue, identifying them as definite, likely or possible group members. The galaxies targeted were generally of lower luminosity than has previously been spectroscopically observed, extending into the dwarf regime (to $M_B \simeq -13 \; \mbox{mag}$). These targets were then matched to the APM catalogue to obtain improved positional coordinates for our observations. To achieve a reasonable signal-to-noise ratio (S/N) in the available observing time, we limited the magnitude of our galaxy targets in these two groups to $b_J \leq 18.5 \; \mbox{mag}$.

The overall effectiveness of our target selection methods is quantified in the yield percentages listed in column (4) of Table \ref{table:GalaxyRedshiftResults}, which represent the percentages of objects with acceptable redshifts that are confirmed group members. These yield percentages show that our target selection method for the Dorado and NGC 1400 groups was clearly more effective, because we only selected targets from the existing group catalogues. The yields for NGC 681, NGC 4038, NGC 4697 and NGC 5084 show what we achieved by selecting targets from a morphological parameter space rather than a pre-defined catalogue.

\subsection{Observing Information}

We used a 300B grating with a central wavelength of 5806 \AA, providing a variety of strong absorption lines for redshift measurement. The targets for each group were split into two magnitude ranges to optimise observing efficiency. Our plan was to achieve a S/N sufficient for strong absorption line recognition and therefore accurate redshift measurement. Despite cloud and technical issues on some nights, we generally achieved exposures with acceptable seeing.

We observed each group in a single two-degree diameter circular field centred on the group position obtained from the NASA Extragalactic Database (NED)\footnote{The NASA/IPAC Extragalactic Database (NED) is operated by the Jet Propulsion Laboratory, California Institute of Technology, under contract with the National Aeronautics and Space Administration. The database is accessible at http://nedwww.ipac.caltech.edu}, which quotes the positions of \citet{Garcia..1993}. Table \ref{table:GalGroupObs} summarises our observations - column (1) is the group name; columns (2) and (3) are the coordinates of the 2dF field centres; columns (4) to (7) provide further observing information.

% Observation Timings/Seeing from document file GroupObsAnalysis_Progress.sxw
% No. of targets from Xmeasured file for each group.
% Target data from .fld files with highest _n (latest) at /net/somerville1/katya/2df/Dorado/
\begin{table*}
\caption{Galaxy Group 2dF Observations}
\centering
\label{table:GalGroupObs}
\begin{tabular*}{1.00\textwidth}
     {@{\extracolsep{\fill}}lcccccc}
\hline
Group Name & RA(J2000) & Dec(J2000) & Date & $b_J$ Magnitude & Exposure & Seeing\\
 & \textit{h:m:s} & \textit{d:m:s} & & Range & \textit{minutes} & \textit{arcsec}\\
(1) & (2) & (3) & (4) & (5) & (6) & (7)\\
\hline
Dorado & 04:17:03.9 & -56:07:43 & 12/11/04 & \textless 14.0 & 120 & 1.2\\
 & & & 13/11/04 & 14.0--18.5 & 90 & 2.2\\
 & & & 13/11/04 & 14.0--18.5 & 90 & 3.2\\
 \\
NGC 681 & 01:49:49.7 & -10:03:05 & 13/11/04 & \textless 13.5 & 120 & 3.0\\
 & & & 14/11/04 & 13.5--18.5 & 90 & 1.2\\
 & & & 14/11/04 & 13.5--18.5 & 60 & 1.5\\
\\
NGC 1400 & 03:40:25.0 & -18:37:16 & 13/11/04 & \textless 12.5 & 120 & 2.4\\
 & & & 14/11/04 & 12.5--18.5 & 180 & 1.2\\
\\
\hline
NGC 4038 & 11:59:57.2 & -19:16:21 & 3/4/05 & \textless 16.5 & 120 & 1.8\\
 & & & 5/4/05 & 16.5--18.5 & 180 & 1.9\\
\\
NGC 4697 & 12:53:55.1 & -06:15:41 & 3/4/05 & \textless 17.1 & 116 & 1.8\\
 & & & 5/4/05 & 17.1--18.5 & 180 & 1.9\\
\\
NGC 5084 & 13:21:30.4 & -21:15:18 & 3/4/05 & $\leq 16.82$ & 53 & 1.5\\
 & & & 3/4/05 & $\leq 16.82$ & 107 & 1.8\\
  & & & 5/4/05 & 16.82--18.5 & 150 & 1.7\\
\hline
\end{tabular*}
\end{table*}

\subsection{Data Reduction and Redshift/Velocity Measurement}

The raw 2dF multi-object spectra images were reduced using 2dfdr\footnote{The 2dF Data Reduction software is available from ftp.aao.gov.au/pub/2df/.} software and IRAF's 1-D spectrum tasks.  We obtained redshifts (converted to recessional velocities) by cross-correlation with a standard set of stellar and galactic templates using the IRAF task XCSAO, following the technique described for the Fornax Cluster redshift survey \citep{Drinkwater..2000}. Our acceptance cut-off for the cross-correlation result is an R-coefficient $\geq 3$ \citep{Tonry..1979}.  We visually inspected and confirmed the redshifts for all member galaxies identified from our observations.  Relatively poor seeing (2.2 to 3.2 arcsec) on some nights (see Table \ref{table:GalGroupObs}) resulted in a lower S/N, consequently reducing the number of confirmed redshifts for these target sets.

\subsection{Group Membership Redefinition}

Table \ref{table:GalaxyRedshiftResults} summarises our galaxy target redshift results -- column (1) is the group name, column (2) is the number of galaxies targeted, column (3) is the number of acceptable galaxy redshifts obtained (R-coefficient $\geq 3$) with bracketed redshift completeness percentage, and column (4) is the number of confirmed group members (see Tables \ref{table:Dorado} to \ref{table:NGC5084}) with bracketed yield (percentage of reliable 2dF redshifts from our observations which are identified as group members).

We obtained galaxy group membership lists from the source catalogues of \citet{Ferguson..1990} and \citet{Garcia..1993}. We supplemented these catalogues with any other objects from our 2dF galaxy and point-source target sets (which extend 1$^{\circ}$ radially from the group centre) whose redshifts place them within the group heliocentric radial velocity limits.

Existing galaxy redshift information for these redefined group catalogues was then extracted from NED, and supplemented by more recent redshift information from the 6dF galaxy redshift survey (6dFGS DR2) issued May 2005 \citep{Jones..2005} and our 2dF observations. The resulting redshift coverage is indicated in Table \ref{table:Redshiftsources}. We also searched the 2dFGRS database \footnote{2dF Galaxy Redshift Survey: http://mcp1.anu.edu.au/~TDFgg/} for redshift information -- only the Dorado field has been covered by this survey and none of the reliable redshifts relate to the group. HIPASS survey \footnote{HI Parkes All-Sky Survey: http://hipass.aus-vo.org/} results are included in NED, but in nearly all cases NED provides alternative redshift sources with lower uncertainty.

% Galaxy-only *** Redshifts from Xmeasured3 files compared to targets from Xmeasured files. Galaxy targets identified by Katya from .fld files.
% Xmeasured3 file temporarily changed to .csv and imported to spreadsheet, sorted by velocity and figures extracted for stars, background, members.
\begin{table}
\caption{Summary of 2dF Redshift Results}
\label{table:GalaxyRedshiftResults}
\begin{tabular}{lccc}\\ 
\hline
Group & Galaxy & Acceptable & Group\\
Name & Targets & Redshifts & Members\\
(1) & (2) & (3) & (4)\\ 
\hline
Dorado & 25 & 18 (72\%) & 8 (44\%)\\
NGC 681 & 41 & 37 (90\%) & 7 (19\%)\\
NGC 1400 & 52 & 46 (88\%) & 29 (63\%)\\
NGC 4038 & 32 & 30 (94\%) & 4 (13\%)\\
NGC 4697 & 55 & 49 (89\%) & 6 (12\%)\\
NGC 5084 & 55 & 45 (82\%) & 10 (22\%)\\
\hline
\end{tabular}
\end{table}

Table \ref{table:Redshiftsources} summarises changes to the original catalogue definitions of the six galaxy groups as a result of subsequent available redshift data from NED, 6dFGS DR2 and our 2dF observations. Column (1) is the group name; columns (2) to (3) show the number of candidate member galaxies listed for this group and the major source catalogues used; columns (4) to (6) show the number of redshift measurements available in NED or 6dFGS DR2, and the number of 2dF redshifts we have obtained; columns (7) and (8) show the changes to the original catalogued group membership as a result of redshift information, being new group members added and previously listed group members excluded (as background galaxies or interlopers); columns (9) and (10) show the revised numbers of redshift-confirmed members and presently unconfirmed members (with no redshift measurement). The figures in Table \ref{table:GalaxyRedshiftResults} (column 3) and Table \ref{table:Redshiftsources} (column 6) agree for Dorado and NGC 1400 (all our 2dF targets came from the original catalogues), but differ for the other four groups (only a proportion of our targets are in the original catalogues).

\begin{table*}
\caption{Summary of Galaxy Group Redefinition - Number of Member Galaxies}
\label{table:Redshiftsources}
\begin{tabular*}{1.00\textwidth}
     {@{\extracolsep{\fill}}lcc|ccc|cc|cc} 
\hline
Group & \multicolumn{2}{c}{Original Catalogues} & \multicolumn{3}{c}{Redshift Data} & \multicolumn{2}{c}{Membership Changes} & \multicolumn{2}{c}{Revised Membership}\\
\cline{2-10}
Name & Members & Source & NED & 6dFGS DR2 & 2dF & Added & Excluded & Confirmed & Unconfirmed\\
(1) & (2) & (3) & (4) & (5) & (6) & (7) & (8) & (9) & (10)\\
\hline
Dorado & 79 & Ferguson \& Sandage & 16 & 7 & 18 & - & 11 & 20 & 48\\
NGC 681 & 4 & Garcia (LGG33) & 7 & - & 8 & 4 & 1 & 7 & -\\
NGC 1400 & 120 & Ferguson \& Sandage & 28 & 13 & 46 & - & 20 & 31 & 69\\
NGC 4038 & 27 & Garcia (LGG263) & 29 & 4 & 5 & 1 & 1 & 27 & -\\
NGC 4697 & 18 & Garcia (LGG314) & 22 & - & 6 & 6 & - & 24 & -\\
NGC 5084 & 5 & Garcia (LGG345) & 10 & 4 & 12 & 10 & 2 & 13 & -\\
\hline
\end{tabular*}
\end{table*}

Tables \ref{table:Dorado} to \ref{table:NGC5084} are detailed listings of group members having redshift information, together with changes to each galaxy group as a result of our observations and analysis. The group member galaxies are in general clearly isolated on the redshift (or heliocentric radial velocity) axis from background galaxies. Where more than one velocity measurement is shown, our approach was to take the velocity with the smallest uncertainty, unless there was a specific reason to do otherwise as noted in the tables. Our dynamical calculations are then based on the revised group membership from these tables. The velocity ranges defining group membership are compared in the heliocentric radial velocity histograms shown in Fig. \ref{fig:groups_vel_hist}.

\section{Analysis}

For convenience, we first present the key kinematical and luminosity-weighted dynamical results of our analysis in Table \ref{table:group_statistics}, based upon the revised group membership and galaxy redshifts. Column (1) is the group name; column (2) shows the number of redshift-confirmed member galaxies; column (3) is the fraction of confirmed members to estimated total members (N), estimated by applying the redshift-confirmed member/non-member ratio to unconfirmed group members; column (4) is the radial velocity dispersion of confirmed group members; column (5) is the distance to the group based on a first-order approximation of the Hubble redshift-distance relation ($H_0$d = cz, where $H_0$ = $72 \; \mbox{km} \, \mbox{s}^{-1} \, \mbox{Mpc}^{-1}$); columns (6) to (8) locate the group centre of mass in redshift space; column (9) is the virial mass estimate; column (10) is the total luminosity of group members; column (11) is the mass-to-light ratio based on dynamical calculations of the group virial mass; column (12) is the harmonic mean radius and column (13) is the group crossing time.

% data comes from running MATLAB on harmeanradius2.m with ...newfile.dat as input for each group
\begin{table*}
\caption{Galaxy Group Kinematical and Dynamical Statistics}
\label{table:group_statistics}
\begin{tabular*}{1.00\textwidth}
     {@{\extracolsep{\fill}}lccccccc} 
\hline
Group & Confirmed & Completeness & Velocity & Distance & \multicolumn{3}{|c|}{Centre of Mass}\\
\cline{6-8}
Name & Members &  & Dispersion & (Modulus) & RA & Dec & cz\\
  & [$n$] & [$n/N$] & [$\mbox{km} \, \mbox{s}^{-1}$] & [Mpc (mag)] & [h:m:s] & [d:m:s] & [$\mbox{km} \, \mbox{s}^{-1}$]\\
(1) & (2) & (3) & (4) & (5) & (6) & (7) & (8)\\
\hline
Dorado & 20 & 0.39 & 222$\,\pm\,$17 & 16.9 (31.1) & 04:17:01.8 & -55:42:46 & 1246$\,\pm\,$39\\
NGC 681 & 7 & 1.00 & 79$\,\pm\,$13 & 25.3 (32.0) & 01:50:04.7 & -10:00:53 & 1815$\,\pm\,$15\\
NGC 1400 & 31 & 0.40 & 207$\,\pm\,$22 & 23.0 (31.8) & 03:42:15.2 & -18:31:04 & 1728$\,\pm\,$23\\
NGC 4038 & 26$^a$ & 1.00 & 75$\,\pm\,$8 & 23.7 (31.9) & 11:59:22.3 & -19:15:18 & 1671$\,\pm\,$11\\
NGC 4697 & 24 & 1.00 & 142$\,\pm\,$9 & 18.9 (31.4) & 12:55:01.6 & -06:24:06 & 1342$\,\pm\,$21\\
NGC 5084 & 13 & 1.00 & 107$\,\pm\,$16 & 25.5 (32.0) & 13:21:53.2 & -21:17:47 & 1763$\,\pm\,$15\\
\\
\hline
\\
\\
\end{tabular*}

\begin{tabular*}{1.00\textwidth}
     {@{\extracolsep{\fill}}lccccc} 
\hline
Group & Virial & Total & M/L & Harmonic & Crossing\\
Name & Mass & Luminosity & Ratio & Radius & Time/$H_0$\\
  & [$10^{13} \; \mbox{M}_{\odot}$] & [$10^{11} \; \mbox{L}_{\odot}$] &  & [Mpc] & \\
(1) & (9) & (10) & (11) & (12) & (13)\\
\hline
Dorado & 1.28$\,\pm\,$0.33 & 0.78$\,\pm\,$0.16 & 171$\,\pm\,$59 & 0.23$\,\pm\,$0.04 & 0.126$\,\pm\,$0.006\\
NGC 681 & 0.17$\,\pm\,$0.05 & 0.10$\,\pm\,$0.03 & 191$\,\pm\,$89 & 0.25$\,\pm\,$0.03 & 0.083$\,\pm\,$0.004\\
NGC 1400 & 1.24$\,\pm\,$0.26 & 0.52$\,\pm\,$0.11 & 257$\,\pm\,$92 & 0.26$\,\pm\,$0.01 & 0.067$\,\pm\,$0.008\\
NGC 4038 & 0.32$\,\pm\,$0.08 & 2.28$\,\pm\,$0.51 & 15$\,\pm\,$6 & 0.51$\,\pm\,$0.04 & 0.577$\,\pm\,$0.024\\
NGC 4697 & 1.23$\,\pm\,$0.16 & 0.69$\,\pm\,$0.13 & 184$\,\pm\,$42 & 0.56$\,\pm\,$0.03 & 0.296$\,\pm\,$0.017\\
NGC 5084 & 0.50$\,\pm\,$0.13 & 0.95$\,\pm\,$0.25 & 59$\,\pm\,$27 & 0.40$\,\pm\,$0.06 & 0.173$\,\pm\,$0.008\\
\\
\hline
\end{tabular*}
\begin{flushleft}
\emph{a.} The colliding galaxies NGC 4038 and NGC 4039 (Antenna galaxies) are treated as a single object in our dynamical calculations, located at their combined, luminosity-weighted centre of mass in redshift space.\\
\end{flushleft}
\end{table*}

In the following sub-sections, we describe the analysis methods, results and interpretation of our findings in detail.

\subsection{Kinematics}

Fig. \ref{fig:groups} shows the relative positions of revised group members, with each group scaled to a common distance based upon the mean group distance shown in Table \ref{table:group_statistics}. This form of presentation allows a direct comparison of group compactness -- Dorado, NGC 4038 and NGC 4697 are loose groups, whereas NGC 681, NGC 1400 and NGC 5084 are compact.

All six groups exhibit multiple peaks in their heliocentric radial velocity plots (Fig. \ref{fig:groups_vel_hist}). The number of galaxies in these groups is not sufficient to interpret these peaks as having physical meaning -- rather they represent the effects of small number statistics. For the same reason, we have not attempted to analyse the shape of these velocity distributions in detail. The velocity dispersions of Dorado and NGC 1400 are higher than the other four groups. Including a proportion of the unconfirmed group members (with no redshift data), these two groups may approach the size of galaxy clusters.  The velocity dispersions of the remaining groups are within the range expected for galaxy groups.

\subsection{Substructure}
% Used Dressler-Shectman Test by converting to MATLAB and selecting approx sqrt(N) = 4 nearest neighbours.

The 1-D (radial velocity histogram) and 2-D (projected sky position) tests for substructure are problematic in the case of galaxy groups, due to the relatively small number of galaxy members. However, the 3-D Dressler-Shectman test for substructure \citep{Dressler..1988}, which was originally designed for cluster size systems, can be adapted to galaxy groups \citep{Zabludoff..1998}.

Briefly, the Dressler-Shectman test identifies the nearest neighbours of each group member galaxy and computes the local mean radial velocity (${v}_{local}$) and velocity dispersion ($\sigma_{local}$) of each resulting subgroup. The squared deviations of the local (compared to the overall group) mean velocities and dispersions are then combined in the $\delta^2$ deviation statistic for each subgroup. This statistic is normalised by a leading term involving the selected number of nearest neighbour galaxies ($nn$) in each subgroup.

\begin{equation}
   \delta^{2} = \left(\frac{nn+1}{\sigma^2}\right)\left[\left(\tilde{v}_{local}-\tilde{v}\right)^2+\left(\sigma_{local}-\sigma\right)^2\right]
   \label{eqn:deviation}
\end{equation}

The sum of the unsquared positive deviations is termed the $\Delta$ statistic. In a randomly distributed system the $\Delta$ statistic is approximately equal to the number of group members with redshift information ($N_v$), whereas $\Delta > N_v$ is an indicator of probable substructure. To test the reliability of the $\Delta$ statistic, a Monte-Carlo analysis is performed -- the velocities are randomly shuffled and the $\Delta$ statistic is recomputed 1000 times, producing the probability P that the observed $\Delta$ is a random result. A low probability is therefore interpreted as evidence of substructure.

After some experimentation, we set the number of nearest neighbours at 4. This is approximately $\sqrt{N_v}$ for our average group as recommended by \citet{Pinkney..1996}, and compares to the figure of 11 nearest neighbours in the original cluster substructure analysis by \citet{Dressler..1988}.

\begin{equation}
   P = \sum\left({\Delta \geq \Delta_{observed}}\right)/{1000}
   \label{eqn:probability}
\end{equation}

\begin{figure*}   \centering
 % fig5a.eps = Dorado_dstest.eps
 % fig5b.eps = NGC681_dstest.eps
 % fig5c.eps = NGC1400_dstest.eps
 % fig5d.eps = NGC4038_dstest.eps
 % fig5e.eps = NGC4697_dstest.eps
 % fig5f.eps = NGC5084_dstest.eps
 \includegraphics[width=0.45\textwidth]{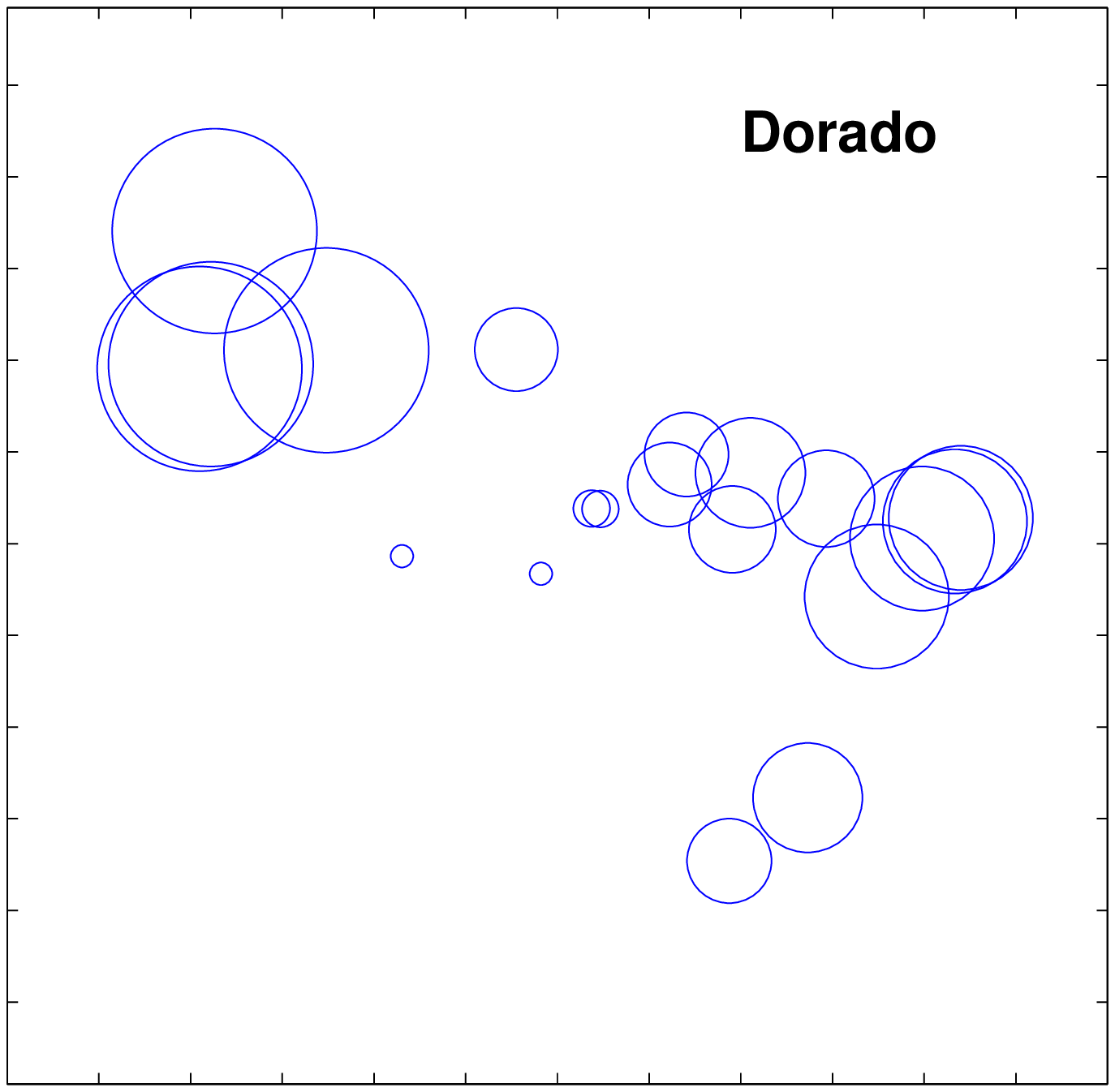}
 \includegraphics[width=0.45\textwidth]{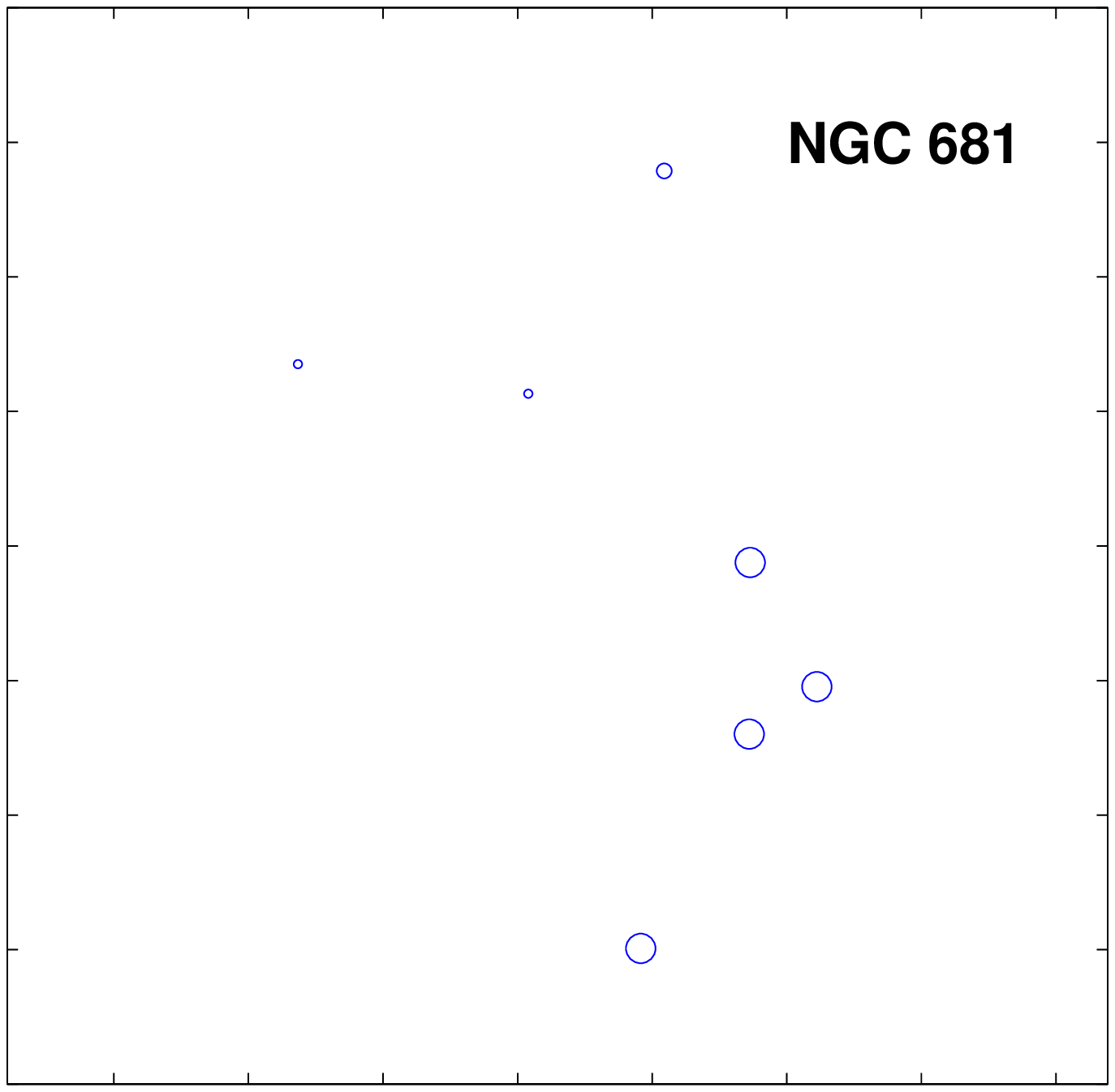}\\
  \includegraphics[width=0.45\textwidth]{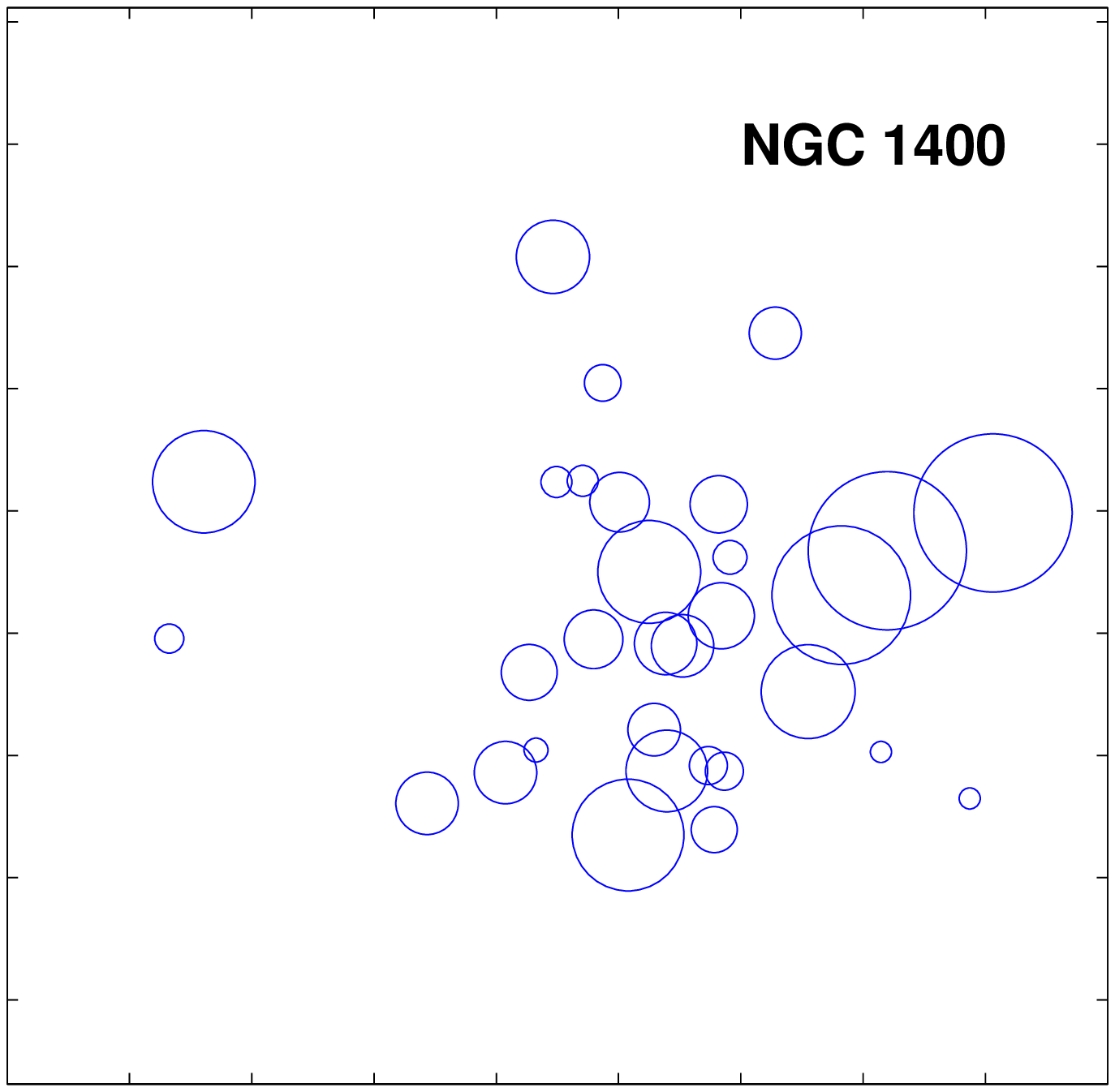}
 \includegraphics[width=0.45\textwidth]{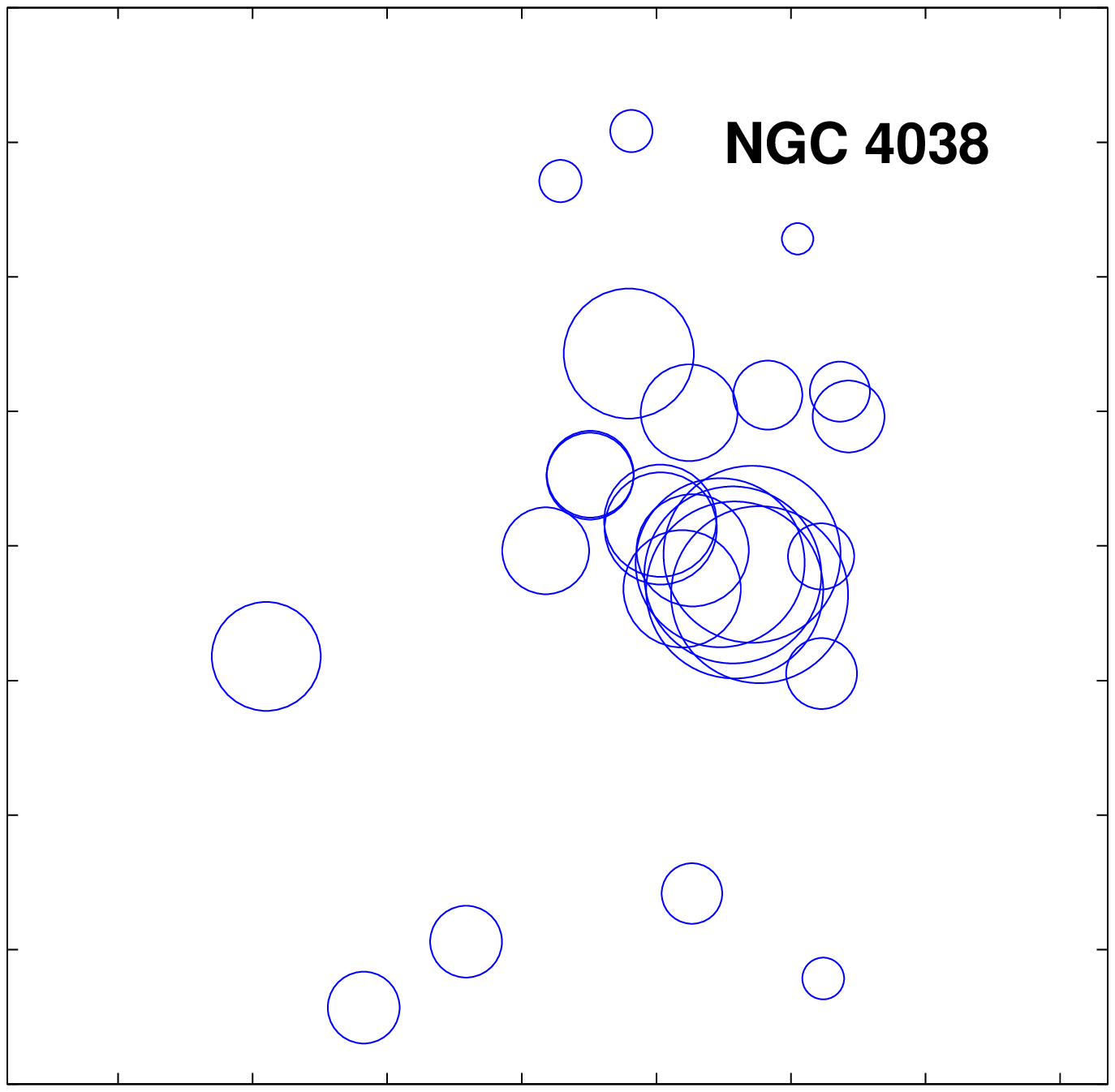}\\
  \includegraphics[width=0.45\textwidth]{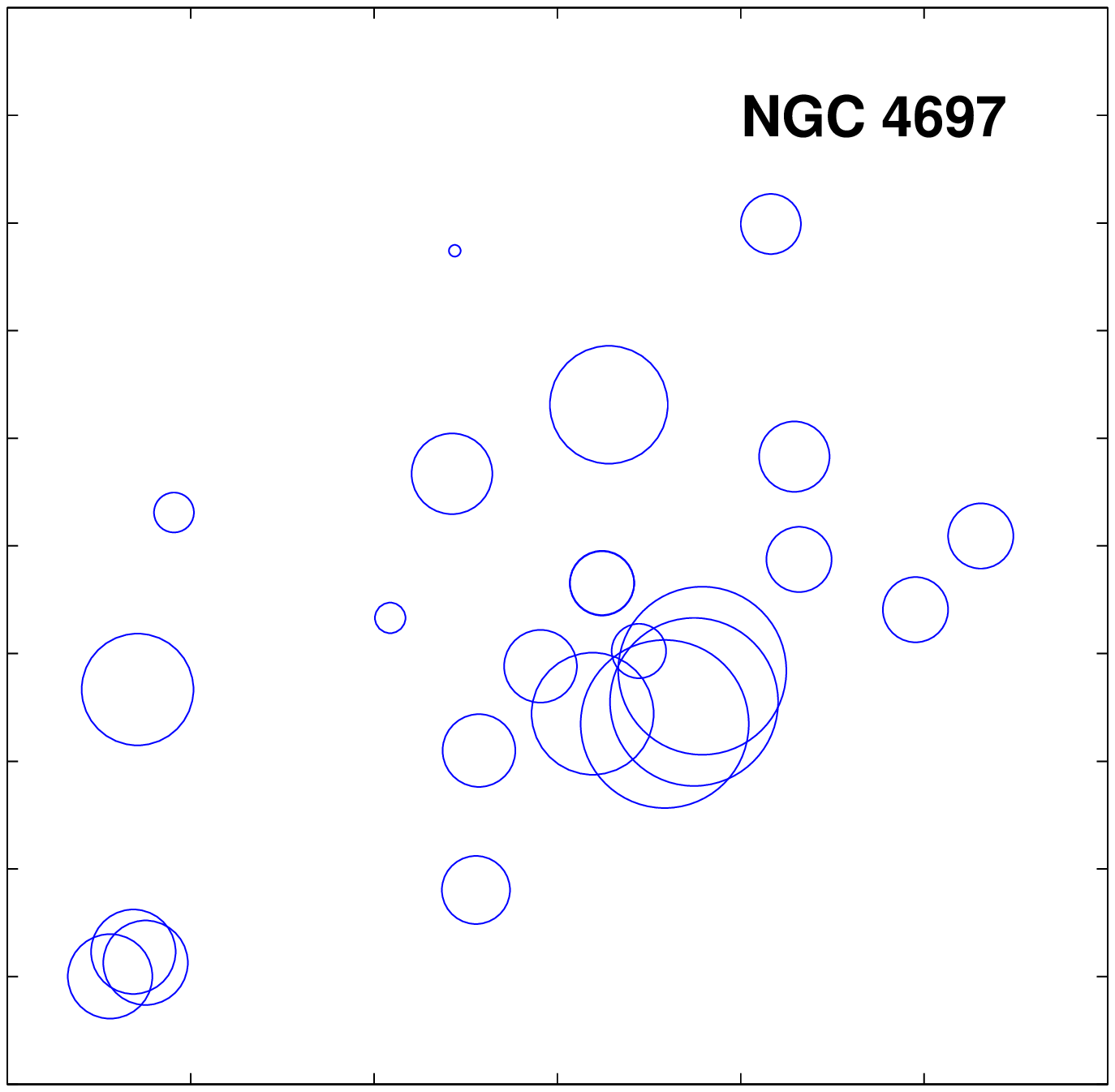}
 \includegraphics[width=0.45\textwidth]{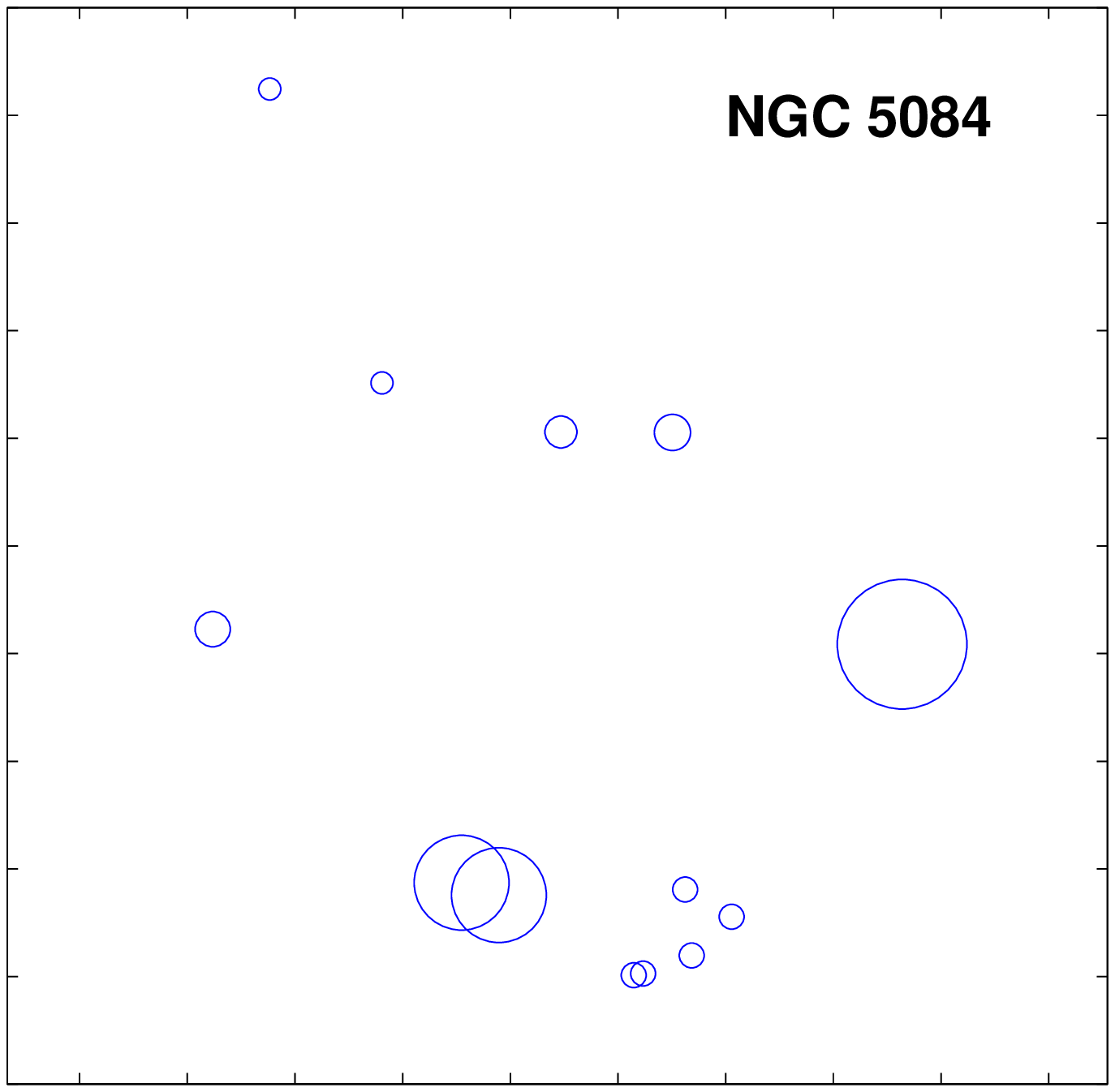}\\
 
 \caption{Bubble plots from the Dressler-Shectman test based on 4 nearest neighbours. The scale of each plot is adjusted to best show the relevant group, so they are not directly comparable. The bubble size is proportional to the squared deviation of the local velocity distribution from the group velocity distribution. A relatively isolated congregation of large bubbles is indicative of a kinematically distinct subgroup. Dorado, NGC 4038 and NGC 4697 show such evidence of substructure.}
 \label{fig:dstest_bubbleplots}
 \end{figure*}

The results of our Dressler-Shectman tests are summarised in Table \ref{table:dstest_results}, and illustrated in the set of bubble plots (Fig. \ref{fig:dstest_bubbleplots}) with the size of bubbles proportional to the $\delta^2$ deviation of each subgroup. The loose groups (Dorado, NGC 4038 and NGC 4697) exhibit sub-structure in redshift space, whereas the compact groups (NGC 681, NGC 1400 and NGC 5084 do not. Without reliable secondary distance indicators, it is unclear whether the sub-groups are at different distances (for example, filamentary structures) or at the same distance but having different recessional velocities (infalling sub-groups).

\begin{table}
\caption{Results of Dressler-Shectman Substructure Test}
\label{table:dstest_results}
\begin{tabular}{lcccc}\\ 
\hline
Group & $N_v$ & $\Delta$ & $\Delta/N_v$ & P\\
(1) & (2) & (3) & (4) & (5)\\ 
\hline
Dorado & 20 & 35.8 & 1.79 & 0.002\\
NGC 681 & 7 & 2.4 & 0.34 & 0.921\\
NGC 1400 & 31 & 32.2 & 1.04 & 0.506\\
NGC 4038 & 27 & 41.8 & 1.55 & 0.044\\
NGC 4697 & 24 & 31.8 & 1.32 & 0.037\\
NGC 5084 & 13 & 9.4 & 0.73 & 0.607\\
\hline
\end{tabular}
\end{table}

\subsection{Luminosity}

The $b_J$ magnitudes for nearly all confirmed group members listed in Tables \ref{table:Dorado} to \ref{table:NGC5084} are from SuperCOSMOS, with magnitude error margins for our luminosity-weighted dynamical calculations based on SuperCOSMOS internal error estimates \citep{Hambly..2001}. Due to limited survey coverage or difficulty in correctly defining the galaxy boundary, SuperCOSMOS gave an unreliable magnitude for a small number of member galaxies -- we substituted SuperCOSMOS bright galaxy measurements (M. Read, private communication) where possible or estimated $b_J$ magnitudes from available NED data (see footnotes in group member lists).

\begin{figure*}   \centering
 \includegraphics[height=10cm, width=0.8\textwidth]{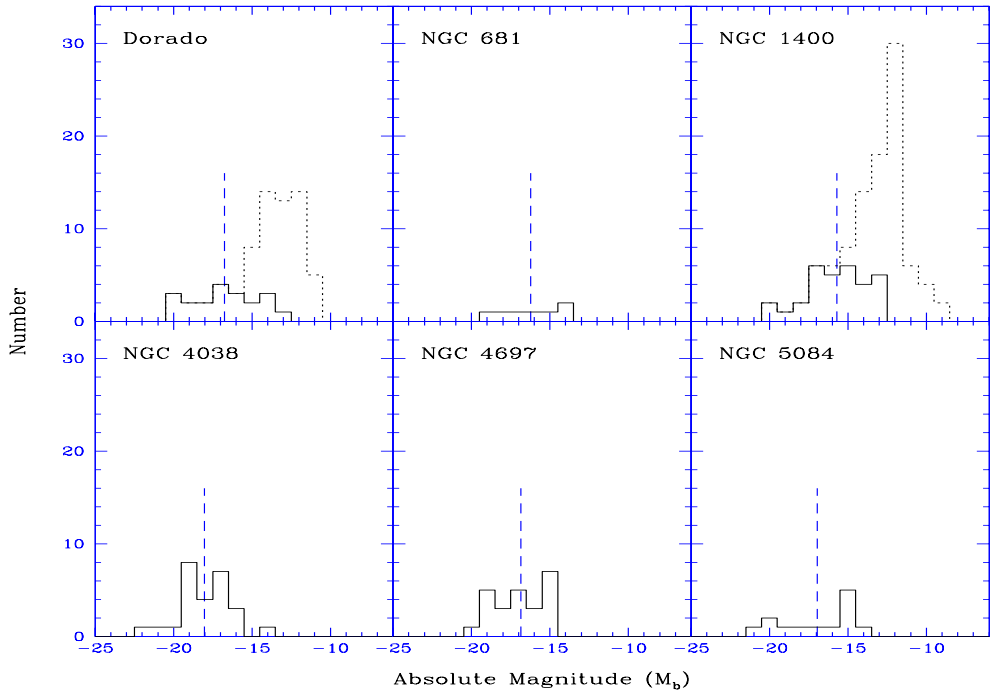}
 \caption{$M_{b_J}$ luminosity functions for redshift-confirmed galaxy group members (solid outlines), with their mean absolute magnitudes (dotted vertical lines). Dorado and NGC 1400 group catalogues contain additional candidate members with no redshift data (dotted outlines - stacked above confirmed members), which are excluded from our dynamical calculations. A proportion of these faint, candidate group member galaxies may be true cluster members, but they will not significantly alter our luminosity-weighted dynamical results.}
 \label{fig:luminosityhist}
 \end{figure*}

Fig. \ref{fig:luminosityhist} depicts the luminosity functions (M$_{b_J}$) of the redshift-confirmed members of each group, in absolute magnitude units based upon distance moduli from Table \ref{table:group_statistics}. The mean absolute magnitude is approximately -17 mag for all groups, except NGC 4038 which is approximately 1 mag brighter due to the starburst activity in the Antennae galaxy pair. For Dorado and NGC 1400 groups, we also show the unconfirmed candidate group members having no redshifts -- we expect that a proportion of these less luminous galaxies will be confirmed as members by future redshift measurements. Our group member lists exclude low surface brightness dwarfs. The impact on our dynamical calculations of excluding unconfirmed candidates and low surface brightness dwarfs is considered negligible, but nevertheless is discussed in the next subsection.

\subsection{Dynamics}
Our group dynamical calculations are luminosity-weighted. The dynamical calculations are based on the formulae summarised in Table \ref{table:formulae}. Our calculations adopt the techniques outlined in \citet{Carlberg..1996}, \citet{Ferguson..1990} and \citet{Materne..1974}. They are similar to Table XIII in the \citet{Ferguson..1990} paper, except as follows: 

\begin{itemize}
\item \emph{Members:} whereas \citet{Ferguson..1990} used all candidate galaxies for position and luminosity dependent calculations, we only use those galaxies with confirmed redshifts for all dynamical calculations.
\item \emph{Weighting:} mean heliocentric radial velocity, velocity dispersion and all mass-related statistics are based on weighting each confirmed member galaxy by its relative apparent $b_J$ magnitude (converted to a relative luminosity).
\item \emph{Error Estimates:} we use Monte Carlo simulations to derive the uncertainties quoted in Table \ref{table:group_statistics}, assuming Gaussian distributions for all input measurements (except the galaxy coordinates which we treat as accurate). We used 100,000 sets of randomised Gaussian input measurements to derive a set of Gaussian output distributions.
\end{itemize}

\begin{table}
\caption{Kinematical and Dynamical Formulae}
\label{table:formulae}
\begin{tabular}{lcr}
\parbox[l]{3cm}{Mean group radial velocity} & \large$\overline{v} = \frac{\sum_iw_iv_i}{\sum_i{w_i}}$ & (3)\\
\\
\parbox[l]{3cm}{Line of sight velocity dispersion} & $\sigma = \left[ \frac{\sum_iw_i(\Delta v_i)^2}{\sum_i{w_i}} \right]^\frac{1}{2}$ & (4)\\
\\
\parbox[l]{3cm}{Projected mean harmonic radius} & $R_H = \left[ \frac{\sum_{i} \sum_{j<i} \frac{w_iw_j}{|r_i-r_j|}}{\sum_{i} \sum_{j<i} w_i w_j} \right]^{-1}$ & (5)\\
\\
\parbox[l]{3cm}{Three-dimensional virial radius} & $r_v = \frac{\pi R_H}{2}$ & (6)\\
\\
\parbox[l]{3cm}{Virial mass} & $M_v = \frac{3}{G} \sigma^2 r_v$ & (7)\\
\\
\parbox[l]{3cm}{Projected mass} & $M_p = \frac{32}{\pi G} \frac{\sum_i w_i (\Delta v_i)^2 r_{\perp i}}{\sum_i w_i}$ & (8)\\ 
\\
\parbox[l]{3cm}{Crossing time} & $t_c = \frac{<r>}{<|v|>}$ & (9)\\
\\
\end{tabular}
\begin{flushleft}
\emph{a.} $\langle r \rangle$ is the average projected distance of group members from the group centre of mass.\\
\emph{b.} $\langle |v| \rangle$ is the average speed of group members relative to the group centre of mass.\\
\emph{c.} Weightings $w_i$ are based on the apparent $b_J$ magnitudes of the group member galaxies, converted to relative luminosities.\\
\emph{d.} G is the gravitational constant.\\
\end{flushleft}
\end{table}

For the NGC 4038 group we treated the colliding galaxy pair (NGC 4038/4039) as a single object with a combined luminosity and located at their centre of mass in redshift space.  This prevents their small and presumably rapidly changing separation from dominating the group and producing an incorrect virial mass estimate. If this is not done, the calculated harmonic mean radius, virial mass and mass-to-light ratio for the group are reduced by an order of magnitude.

The virial mass, total luminosity and consequently the M/L ratio have significant error margins which are partly due to uncertainty in the $b_J$ magnitudes for some of the brighter group members. This uncertainty stems from the difficulty SuperCOSMOS has in defining a reliable area over which to integrate the light from the larger, brighter galaxies. We have mitigated this effect by substituting SuperCOSMOS luminosities optimised for bright galaxies (M. Read, private communication).

\begin{figure*}   \centering
 \includegraphics[height=1.2\textwidth, width=0.8\textwidth]{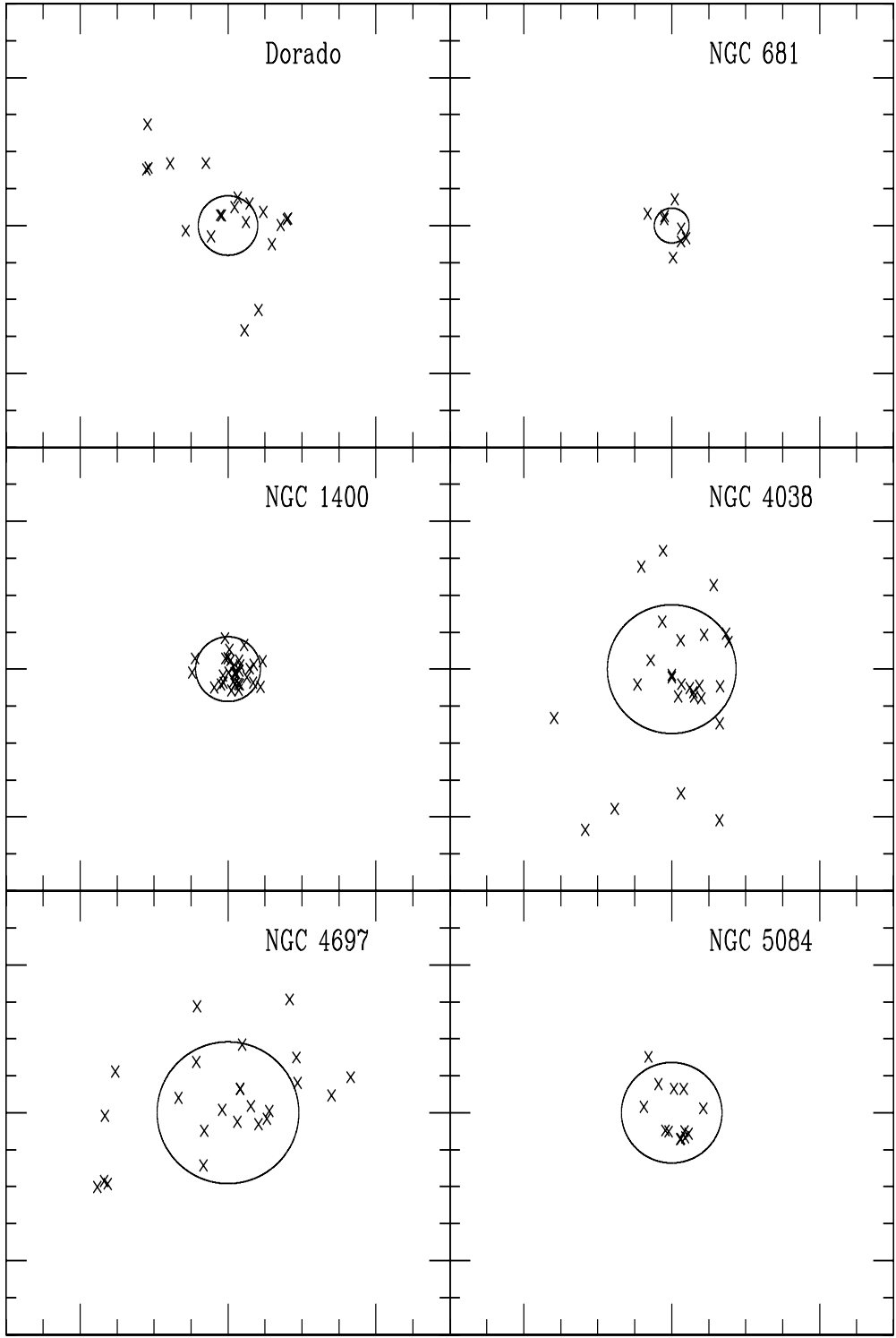}
  \caption{Six galaxy groups showing revised members (i.e. with redshifts) scaled as they would appear if viewed from a common distance. Each square plot would be approximately 3.5 Mpc by 3.5 Mpc if viewed from a common distance of 20 Mpc. The circle represents the luminosity-weighted harmonic mean radius centred on the group centre of mass. The compact and loose groups become obvious in this style of presentation. NGC 4038 and NGC 4697 have large luminosity-weighted harmonic mean radii and relatively long crossing times.}
 \label{fig:groups}
 \end{figure*}

The luminosity-weighted harmonic mean radius of each group is illustrated in Figure \ref{fig:groups}. The position and size of the circles are determined by the group centre of mass and galactic mass distribution respectively. We have not attempted to quantify or map the distribution of inter-galactic matter, either baryonic or non-baryonic. Groups NGC 4038 and NGC 4697 have relatively large luminosity-weighted harmonic mean radii compared to the other groups, reflecting the more widespread projected spatial dispersion of some brighter member galaxies. The loose groups in general have larger harmonic mean radii than the compact groups, although Dorado is an exception with a relatively small harmonic mean radius due to the concentration of more luminous galaxies at its core.

The mass of confirmed group members can be computed in two dynamically-derived ways, following the discussion in \citet{Heisler..1985}. The virial mass ($M_v$) and projected mass estimate ($M_p$), whose definitions are shown in Table \ref{table:formulae}, assume spherical symmetry and an isotropic distribution of velocities.  Heisler et al. concluded that the projected mass estimate was less affected by interlopers than the virial mass estimate, but as we have specifically excluded such interlopers in our calculations the two mass estimates should be equally valid. We have therefore computed virial mass estimates.

The mass-to-light ratios (M/L) have been computed from the ratio of the group virial mass estimate to the sum of group member galaxy luminosities, based upon the use of a group distance modulus to convert apparent $b_J$ magnitudes to absolute magnitudes and then to a total luminosity for all group members. The presence of a significant amount of dark matter is evident from the generally high M/L calculated from our revised group member lists, except for the NGC4038 group where the colliding central galaxy pair (NGC 4038 and NGC 4039) have temporarily boosted star formation activity and total luminosity. NGC 5084 has an intermediate M/L but no dominant starburst galaxies.

For all our groups, we estimate that the impact on our luminosity-weighted results of missing light due to incompleteness in redshift data (unconfirmed catalogue members and uncatalogued dwarf galaxies) is negligible. For example, based upon available redshift data, approximately half of the unconfirmed galaxies in the Dorado (48 unconfirmed) and NGC 1400 (69 unconfirmed) catalogues are likely to be group members, but their estimated missing light is only 0.3\% to 0.4\% of the total luminosity of the confirmed group members. Moreover, the radial distribution of missing light should follow that of the confirmed member galaxies -- for example, \citet{Carrasco..2001} investigated the dwarf galaxy population of the Dorado group, finding that they tended to cluster around the brighter galaxy members. Therefore, particularly after luminosity weighting, missing dwarf galaxies will not significantly alter our group velocity dispersions, virial mass estimates or crossing times. 

The group crossing time is usually compared to the Hubble time to determine whether the groups are virialised \citep{Ferguson..1990}. The derived crossing times for NGC 4038 and NGC 4697 exceed 0.2 Hubble times, suggesting that they are unvirialised, whereas Dorado, NGC 681, NGC 1400 and NGC 5084 appear to be dynamically relaxed. \citet{Ferguson..1990} also found that Dorado and NGC 1400 were virialised. A refinement of this test is to scale the number of crossings required to achieve dynamical relaxation by the estimated number of group member galaxies (N). The relaxation time is then proportional to N/log(0.4N) \citep{Aarseth..1972}. Applying this approach, we would conclude that NGC 681, NGC 1400 and NGC 5084 are dynamically relaxed, whereas Dorado, NGC 4038 and NGC 4697 are not yet virialised. Virial mass estimates for the latter three galaxy groups may therefore be inappropriate.

\begin{figure*}   \centering
 \includegraphics[height=10cm, width=0.8\textwidth]{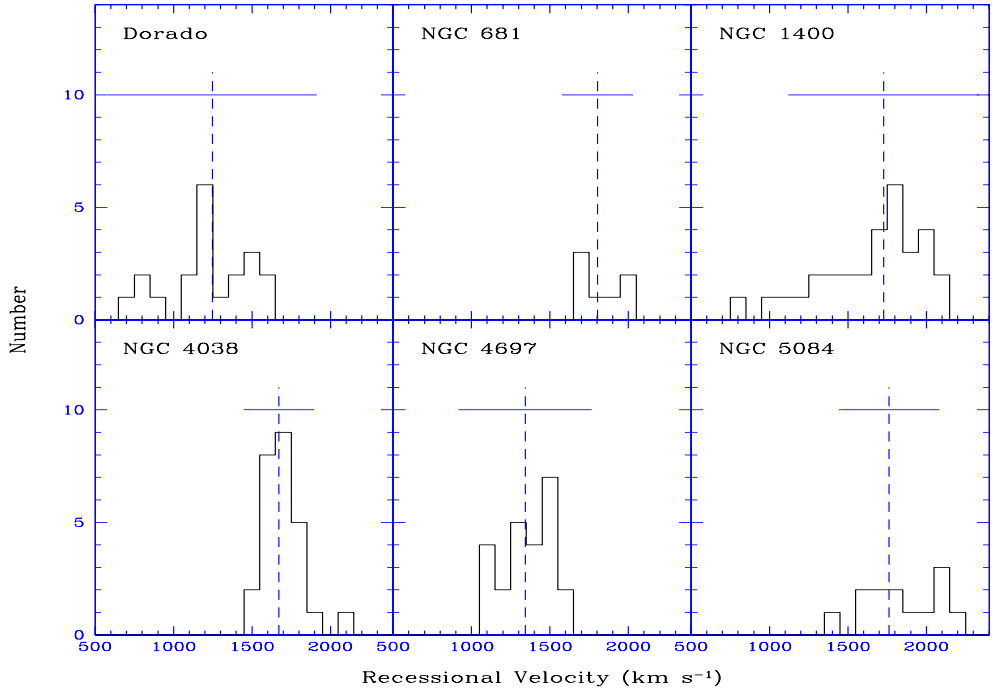}
 \caption{The heliocentric radial velocity distribution of confirmed galaxy group members from our 2dF observations. The luminosity-weighted mean velocity (vertical dotted line) and $3\sigma$ velocity dispersion (horizontal line) illustrate the approximate dynamical boundaries of each group along the radial velocity axis.}
 \label{fig:groups_vel_hist}
 \end{figure*}

The interpretation of apparently relaxed groups is also problematic -- numerical simulations indicate that projection effects and incomplete redshift data for group members can significantly effect dynamical mass estimates, such that even groups with short apparent crossing times may still be in the pre-virial collapse phase \citep{Aarseth..1972,Danese..1981,Diaferio..1993}. However, our luminosity-weighted approach should have largely overcome redshift data deficiencies since, as illustrated in the radial velocity histograms (Figure \ref{fig:groups_vel_hist}), we have redshift data for the dynamically dominant group members.

The virial state of NGC 1400 group depends upon the inclusion or exclusion of galaxy NGC 1400, which has a large radial velocity difference from other group members. There is ample evidence that this galaxy is not a foreground object, but is rather located at the group distance \citep{Gould..1993,Quintana..1994,Perrett..1997}. In our analysis, the galaxy NGC 1400 is treated as an interloper (see Table \ref{table:NGC1400}) and excluded from the dynamical calculations. Our resulting luminosity-weighted group virial mass agrees with \citet{Quintana..1994} under this assumption, and the group is virialised. If we include NGC 1400 in our calculations, the group crossing time is reduced from 0.067 to 0.047 and the mass-to-light ratio is increased from 257$\,\pm\,$92 to 835$\,\pm\,$253 -- again broadly in line with \citet{Quintana..1994} who interprets this as evidence of substantial dark matter. In this case, the virial formulae may not apply if the high relative velocity of NGC 1400 is interpreted as an infalling sub-group. Since there are no other bright galaxies in this group with recessional velocities similar to NGC 1400, we consider it to be an isolated interloper and treat the group as virialised.

\section{Summary of Findings}

Our key findings are as follows:

\begin{itemize}
\item Target selection by morphological parameters (image area and peak intensity against magnitude) has resulted in 12 to 22 per cent of the successfully measured redshifts being identified as group member galaxies.
\item We have redefined the non-dwarf galaxy membership of the six groups with more accurate redshift--velocity measurements, eliminating several background galaxies. However, significant numbers of less luminous galaxies in Dorado and NGC 1400 groups still require redshift measurements.
\item Several techniques confirm that the six galaxy groups can be separated into loose group (Dorado, NGC 4038 and NGC 4697) and compact group (NGC 681, NGC 1400  and NGC 5084) subsets in both kinematical and dynamical terms. Kinematical techniques include the comparison of sky position plots converted to a common group distance, and the Dressler-Shectman substructure test. Dynamical tests (luminosity-weighted) include comparison of harmonic mean radii at a common group distance, and testing for virialisation through the crossing time.
\item We find substructure to be evident from the Dressler-Shectman test in Dorado, NGC 4038 and NGC 4697. A more refined application of the crossing time test for virialisation also confirms these three groups to be still undergoing dynamical relaxation. The luminosity-weighted virial mass estimate and mass-to-light ratio for these groups may therefore be questionable.
\end{itemize}

\section{Acknowledgements}

Our observations were made using the 2dF spectrograph at the Anglo--Australian Observatory. We wish to thank AAO staff for their support and encouragement.

The 6dF Galaxy Survey (Data Release 2) is provided through the Royal Observatory, Edinburgh and the Anglo--Australian Observatory.

The original code for the Dressler-Shectman test was provided by Kevin Pimbblet of the University of Queensland Astrophysics Group. Peter Firth rewrote this code in MATLAB.

We wish to thank Mike Read for access to his bright galaxy magnitude estimates, and the anonymous reviewer for constructive feedback which has improved the final paper.

\bibliography{galgroupdynamics_MNRAS}

\begin{thebibliography}{}

\bibitem[\protect\citeauthoryear{Aarseth \& Saslaw}{Aarseth \&
  Saslaw}{1972}]{Aarseth..1972}
Aarseth S.~J.,  Saslaw W.~C.,  1972, ApJ, 172, 17

\bibitem[\protect\citeauthoryear{Carlberg et~al.,}{Carlberg
  et~al.}{1996}]{Carlberg..1996}
Carlberg R.~G.,  et~al., 1996, ApJ, 462, 32

\bibitem[\protect\citeauthoryear{Carrasco et~al.,}{Carrasco
  et~al.}{2001}]{Carrasco..2001}
Carrasco E.~R.,  et~al., 2001, AJ, 121, 148

\bibitem[\protect\citeauthoryear{Danese, de Zotti et~al.,}{Danese
  et~al.}{1981}]{Danese..1981}
Danese L.,  de Zotti G.,    et~al., 1981, ApJ, 244, 777

\bibitem[\protect\citeauthoryear{de
  Vaucouleurs}{de~Vaucouleurs}{1975}]{deVaucouleurs..1975}
de Vaucouleurs G.,  1975, Stars and Stellar Systems, 9, 557

\bibitem[\protect\citeauthoryear{Diaferio, Ramella, Geller \& Ferrari}{Diaferio
  et~al.}{1993}]{Diaferio..1993}
Diaferio A.,  Ramella M.,  Geller M.~J.,    Ferrari A.,  1993, AJ, 105, 2035

\bibitem[\protect\citeauthoryear{Dressler \& Shectman}{Dressler \&
  Shectman}{1988}]{Dressler..1988}
Dressler A.,  Shectman S.~A.,  1988, AJ, 95, 985

\bibitem[\protect\citeauthoryear{Drinkwater et~al.,}{Drinkwater
  et~al.}{2000}]{Drinkwater..2000}
Drinkwater M.~J.,  et~al., 2000, A\&A, 355, 900

\bibitem[\protect\citeauthoryear{Ferguson \& Sandage}{Ferguson \&
  Sandage}{1990}]{Ferguson..1990}
Ferguson H.~C.,  Sandage A.,  1990, AJ, 100, 1

\bibitem[\protect\citeauthoryear{Garcia}{Garcia}{1993}]{Garcia..1993}
Garcia A.~M.,  1993, A\&A.Supp., 100, 47

\bibitem[\protect\citeauthoryear{Gould}{Gould}{1993}]{Gould..1993}
Gould A.,  1993, ApJ, 403, 37

\bibitem[\protect\citeauthoryear{Hambly, Irwin \& MacGillvray}{Hambly
  et~al.}{2001}]{Hambly..2001}
Hambly N.~C.,  Irwin M.~J.,    MacGillvray H.~T.,  2001, MNRAS, 326, 1295

\bibitem[\protect\citeauthoryear{Heisler, Tremaine \& Bahcall}{Heisler
  et~al.}{1985}]{Heisler..1985}
Heisler J.,  Tremaine S.,    Bahcall J.~N.,  1985, ApJ, 298, 8

\bibitem[\protect\citeauthoryear{Huchra \& Geller}{Huchra \&
  Geller}{1982}]{Huchra..1982}
Huchra J.~P.,  Geller M.~J.,  1982, ApJ, 257, 423

\bibitem[\protect\citeauthoryear{Jones, Saunders, Read \& Colless}{Jones
  et~al.}{2005}]{Jones..2005}
Jones D.~H.,  Saunders W.,  Read M.~A.,    Colless M.,  2005, PASA, 22, 277

\bibitem[\protect\citeauthoryear{{Kregel}, {van der Kruit} \& {de
  Blok}}{{Kregel} et~al.}{2004}]{Kregel..2004}
{Kregel} M.,  {van der Kruit} P.~C.,    {de Blok} W.~J.~G.,  2004, MNRAS, 352,
  768

\bibitem[\protect\citeauthoryear{{Lauberts} \& {Valentijn}}{{Lauberts} \&
  {Valentijn}}{1989}]{1989spce.book.....L}
{Lauberts} A.,  {Valentijn} E.~A.,  1989, {The surface photometry catalogue of
  the ESO-Uppsala galaxies}.
Garching: European Southern Observatory

\bibitem[\protect\citeauthoryear{Maddox, Sutherland, Efstathiou \&
  Loveday}{Maddox et~al.}{1990}]{Maddox..1990}
Maddox S.~J.,  Sutherland W.~J.,  Efstathiou G.,    Loveday J.,  1990, MNRAS,
  243, 692M

\bibitem[\protect\citeauthoryear{Maia, da Costa \& Latham}{Maia
  et~al.}{1989}]{Maia..1989}
Maia M. A.~G.,  da Costa L.~N.,    Latham D.~W.,  1989, ApJ.Supp., 69, 809

\bibitem[\protect\citeauthoryear{Materne \& Tammann}{Materne \&
  Tammann}{1974}]{Materne..1974}
Materne J.,  Tammann G.~A.,  1974, A\&A, 37, 383

\bibitem[\protect\citeauthoryear{Morshidi-Esslinger, Davies \&
  Smith}{Morshidi-Esslinger et~al.}{1999}]{Morshidi..1999}
Morshidi-Esslinger Z.,  Davies J.~I.,    Smith R.~M.,  1999, MNRAS, 304, 297

\bibitem[\protect\citeauthoryear{Perrett, Hanes, Butterworth, Kavelaars,
  Geisler \& Harris}{Perrett et~al.}{1997}]{Perrett..1997}
Perrett K.,  Hanes D.,  Butterworth S.,  Kavelaars J.,  Geisler D.,    Harris
  W.,  1997, AJ, 113, 895

\bibitem[\protect\citeauthoryear{Pinkney, Roettiger \& Burns}{Pinkney
  et~al.}{1996}]{Pinkney..1996}
Pinkney J.,  Roettiger K.,    Burns J.~O.,  1996, ApJ Supp., 104, 1

\bibitem[\protect\citeauthoryear{Quintana, Fouqu\'e \& Way}{Quintana
  et~al.}{1994}]{Quintana..1994}
Quintana H.,  Fouqu\'e P.,    Way M.~J.,  1994, A\&A, 283, 722

\bibitem[\protect\citeauthoryear{Shakhbazian}{Shakhbazian}{1957}]{Shakhbazian.%
.1957}
Shakhbazian R.~K.,  1957, Astron. Tsirk., 177, 11

\bibitem[\protect\citeauthoryear{Tonry \& Davis}{Tonry \&
  Davis}{1979}]{Tonry..1979}
Tonry J.,  Davis M.,  1979, AJ, 84, 1511

\bibitem[\protect\citeauthoryear{Toomre \& Toomre}{Toomre \&
  Toomre}{1972}]{Toomre..1972}
Toomre A.,  Toomre J.,  1972, ApJ, 178, 623

\bibitem[\protect\citeauthoryear{Tully}{Tully}{1987}]{Tully..1987}
Tully R.~B.,  1987, ApJ, 321, 280

\bibitem[\protect\citeauthoryear{Zabludoff \& Mulchaey}{Zabludoff \&
  Mulchaey}{1998}]{Zabludoff..1998}
Zabludoff A.~I.,  Mulchaey J.~S.,  1998, ApJ, 498, L5

\end{thebibliography}

% fig2.eps = groups_samedistplot.eps (input groups_samedistplot.sm)

 % fig3.eps = groups_velhist2.eps (input groups_vel_hist2.sm)

 % fig4.eps = groups_luminosityhist.eps (input groups_luminosityhist.sm)

 % MATLAB DS-test eps files from ds_test.m 

\begin{table*}
\caption{Dorado Group}
\label{table:Dorado}
\begin{tabular*}{1.00\textwidth}
     {@{\extracolsep{\fill}}llccccccc} 
\hline
Ref. & Name & RA(J2000) & Dec(J2000) & $b_{j}$ & \multicolumn{3}{|c|}{Velocity ($\mbox{km} \, \mbox{s}^{-1})$} & Revised\\
\cline{6-8}
 &  & [h:m:s] & [d:m:s] &  & NED & 6dF DR2 & 2dF Data & Status\\
(1) & (2) & (3) & (4) & (5) & (6) & (7) & (8) & (9)\\ 
\hline
FS07 & AM 0407-553 & 04:08:53.9 & -55:27:33 & 16.26 & 29673$\,\pm\,$46 & - & - & background\\
FS08 & IC 2038 & 04:08:53.7 &  -55:59:22 & 15.17 & 712$\,\pm\,$52 & - & - & 712$\,\pm\,$52\\
FS12 & IC 2039 $^a$ & 04:09:02.4 &  -56:00:42 & 15.23 & 250$\,\pm\,$100 & 857$\,\pm\,$46 & - & 857$\,\pm\,$46\\
FS16 & NGC 1533 & 04:09:51.8 &  -56:07:06 & 12.47 & 785$\,\pm\,$8 & 764$\,\pm\,$46 & - & 785$\,\pm\,$8\\
FS21 & J04104956-5621105 & 04:10:49.6 &  -56:21:10 & 16.41 & - & - & 8593$\,\pm\,$14 & background\\
FS22 & NGC 1536 & 04:10:59.8 &  -56:28:50 & 13.61 & 1296$\,\pm\,$28 & 1274$\,\pm\,$46 & 1214$\,\pm\,$13 & 1217$\,\pm\,$13\\
FS24 & J04115703-5647245 & 04:11:57.0 &  -56:47:25 & 17.19 & - & - & 26778$\,\pm\,$19 & background\\
FS26 & LSBG $^d$ F157-059 & 04:12:15.5 & -55:52:07 & 17.57 & - & - & 1216$\,\pm\,$125 & 1216$\,\pm\,$125\\
FS28 & NGC 1543 & 04:12:43.2 & -57:44:17 & 12.00 & 1088$\,\pm\,$24 & 1149$\,\pm\,$46 & - & 1088$\,\pm\,$24\\
FS30 & LSBG F157-057 & 04:14:09.1 & -55:42:27 & 16.84 & - & - & 1249$\,\pm\,$81 & 1249$\,\pm\,$81\\
FS31 & NGC 1546 & 04:14:36.5 & -56:03:39 & 13.06 & 1276$\,\pm\,$28 & 1238$\,\pm\,$46 & 1161$\,\pm\,$17 & 1161$\,\pm\,$17\\
FS33 & 6dF J0414409-580755 & 04:14:40.9 & -58:07:55 & 16.52 & 1242$\,\pm\,$46 & - & - & 1242$\,\pm\,$46\\
FS35 & NGC 1549 & 04:15:45.1 & -55:35:32 & 11.39 & 1220$\,\pm\,$15 & 1202$\,\pm\,$46 & - & 1220$\,\pm\,$15\\
FS37 & NGC 1553 & 04:16:10.5 & -55:46:49 & 11.20 & 1080$\,\pm\,$11 & 1172$\,\pm\,$46 & - & 1080$\,\pm\,$11\\
FS39 & IC 2058 & 04:17:54.3 & -55:55:58 & 14.17 & 1379$\,\pm\,$1 & - & 1309$\,\pm\,$17 & 1379$\,\pm\,$1 $^b$\\
FS41 & J04180709-5555503 & 04:18:07.1 & -55:55:50 & 16.25 & - & - & 1369$\,\pm\,$25 & 1369$\,\pm\,$25\\
FS44 & B041735.16-565655.9 & 04:18:36.4 & -56:49:45 & 17.60 & - & - & 22975$\,\pm\,$55 & background\\
FS45 & B041748.01-560337.6 & 04:18:52.2 & -55:56:27 & 17.63 & - & - & 32275$\,\pm\,$28 & background\\
FS46 & B041800.03-554847.5 & 04:19:05.0 & -55:41:38 & 17.05 & - & - & 26069$\,\pm\,$8 & background\\
FS49 & LSBG F157-089 & 04:19:23.4 & -56:20:17 & 17.66 & - & - & 810$\,\pm\,$77 & 810$\,\pm\,$77\\
FS51 & LSBG F157-055 & 04:19:32.9 & -55:29:01 & 18.31 & - & - & 24875$\,\pm\,$30 & background\\
FS53 & NGC 1566 & 04:20:00.4 & -54:56:16 & 10.92 & 1504$\,\pm\,$2 & - & - & 1504$\,\pm\,$2\\
FS56 & B042039.38-555149.1 & 04:21:43.7 & -55:44:50 & 17.33 & - & - & 10347$\,\pm\,$13 & background\\
FS57 & J04214535-5637227 & 04:21:45.4 & -56:37:23 & 17.77 & - & - & 13003$\,\pm\,$25 & background\\
FS58 & B042051.37-563217.1 $^c$ & 04:21:53.4 & -56:25:19 & 19.57 & - & - & 4851$\,\pm\,$22 & background\\
FS59 & J04220378-5621278 & 04:22:03.8 & -56:21:28 & 16.55 & - & - & 11320$\,\pm\,$56 & background\\
FS61 & APMBGC 157+016+068 & 04:22:51.7 & -56:13:39 & 17.15 & 1350$\,\pm\,$4 & - & 1326$\,\pm\,$115 & 1350$\,\pm\,$4\\
FS64 & NGC 1581 & 04:24:44.9 & -54:56:31 & 13.86 & 1600$\,\pm\,$27 & - & - & 1600$\,\pm\,$27\\
FS75 & ESO 157-G 030 & 04:27:32.6 & -54:11:48 & 15.06$^e$ & 1471$\,\pm\,$28 & - & - & 1471$\,\pm\,$28\\
FS76 & NGC 1596 & 04:27:38.1 & -55:01:40 & 12.94 & 1510$\,\pm\,$8 & - & - & 1510$\,\pm\,$8\\
FS78 & NGC 1602 & 04:27:55.0 & -55:03:28 & 14.04 & 1568$\,\pm\,$8 & - & - & 1568$\,\pm\,$8\\
\\
  & \multicolumn{7}{l}{plus 48 other Ferguson \& Sandage catalogue members with no redshift information}\\
\hline
\end{tabular*}
\begin{flushleft}
\emph{a.} The NED velocity is an based on a distance estimate from surface brightness properties \citep{1989spce.book.....L}. The revised 6dF DR2 velocity has the highest level of confidence (quality 4) and is assumed to supercede the previous result.\\
\emph{b.} This is an edge-on spiral galaxy (ESO157-G18) with a velocity range of $224 \; \mbox{km} \, \mbox{s}^{-1})$ measured across its disk due to rotation \citep{Kregel..2004}. The NED velocity is the central velocity, whereas our observations may have targeted an approaching part of the disk.\\
\emph{c.} This target was selected as a point source (UCD candidate) rather than as a galaxy.\\
\emph{d.} Low surface brightness galaxies (LSBG) in these tables are from \citet{Morshidi..1999}.\\
\emph{e.} $b_J$ magnitude based on APM galaxy survey data \citep{Maddox..1990}.\\
\end{flushleft}
\end{table*}

\begin{table*}
\caption{NGC 681 Group (LGG 33)}
\label{table:NGC681}
\begin{tabular*}{1.00\textwidth}%
     {@{\extracolsep{\fill}}llccccccc} 
\hline
Ref. & Name & RA(J2000) & Dec(J2000) & $b_{j}$ & \multicolumn{2}{|c|}{Velocity ($\mbox{km} \, \mbox{s}^{-1})$} & Revised\\
\cline{6-7}
 &  & [h:m:s] & [d:m:s] &  & NED & 2dF Data & Status\\
(1) & (2) & (3) & (4) & (5) & (6) & (7) & (8)\\ 
% 14 km/s subtracted from 2dF Xmeasured3 values to correct to heliocentric velocities
\hline
PGC6667 & MCG -02-05-051 & 01:48:36.4 & -10:19:35 & 15.59 & 1588$\,\pm\,$18 & 1692$\,\pm\,$66 & 1692$\,\pm\,$66 $^a$\\
PGC6671 & NGC 681 & 01:49:10.9 & -10:25:38 & 13.85 & 1745$\,\pm\,$9 & 1857$\,\pm\,$157 & 1745$\,\pm\,$9\\
PGC6826 & NGC 701 & 01:51:03.8 & -09:42:10 & 13.38 & 1831$\,\pm\,$5 & 1839$\,\pm\,$8 & 1839$\,\pm\,$8\\
PGC6832 & IC 1738 & 01:51:08.1 & -09:47:30 & 15.52$^c$ & 1750$\,\pm\,$17 & 12946$\,\pm\,$29 & background $^b$\\
- & UGCA 021 & 01:49:10.4 & -10:03:44 & 15.04 & 1985$\,\pm\,$3 & 2070$\,\pm\,$22 & 1985$\,\pm\,$3\\
- & SDSS J014954.29-091341.2 & 01:49:54.3 & -09:13:42 & 17.20 & 1856$\,\pm\,$43 & 2055$\,\pm\,$89 & 1856$\,\pm\,$43\\
- & B014738.45-110753.6 & 01:50:06.3 & -10:53:02 & 18.26 & - & 1956$\,\pm\,$83 & 1956$\,\pm\,$83\\
- & SDSS J015301.34-093822.1 & 01:53:01.4 & -09:38:22 & 18.43 & 1670$\,\pm\,$20 & 1658$\,\pm\,$16 & 1658$\,\pm\,$16\\
\\
\hline
\end{tabular*}
\begin{flushleft}
\emph{a.} Two velocity measurements are quoted in NED: 1588$\,\pm\,$18 and 1906$\,\pm\,$98. Our 2dF result lies between these.\\
\emph{b.} The NED velocity is from a 21-cm Hydrogen line measurement in the 3rd Reference Catalogue of Bright Galaxies by de Vaucoleurs.\\
\emph{c.} $b_J$ magnitude from SuperCOSMOS bright galaxy measurements (M. Read, private communication).\\
\end{flushleft}
\end{table*}

\begin{table*}
\caption{NGC 1400 Group}
\label{table:NGC1400}
\begin{tabular*}{1.00\textwidth}%
     {@{\extracolsep{\fill}}llccccccc} 
\hline
Ref. & Name & RA(J2000) & Dec(J2000) & $b_{j}$ & \multicolumn{3}{|c|}{Velocity ($\mbox{km} \, \mbox{s}^{-1})$} & Revised\\
\cline{6-8}
 &  & [h:m:s] & [d:m:s] &  & NED & 6dF DR2 & 2dF Data & Status\\
(1) & (2) & (3) & (4) & (5) & (6) & (7) & (8) & (9)\\ 
\hline
FS001 & J03360547-1913359 & 03:36:05.5 & -19:13:36 & 17.18 & 7488$\,\pm\,$46 & - & - & background\\
FS004 & J03361384-1804039 & 03:36:13.8 & -18:04:04 & 16.69 & 23539$\,\pm\,$46 & - & - & background\\
FS010 & NGC 1383 & 03:37:39.2 & -18:20:22 & 14.14 & 1948$\,\pm\,$19 & 2008$\,\pm\,$46 & 2032$\,\pm\,$52 & 2008$\,\pm\,$46 $^c$\\
FS011 & B033525.26-182614.3 & 03:37:41.0 & -18:16:27 & 18.27 & - & - & 18538$\,\pm\,$23 & background\\
FS013 & NGC 1390 & 03:37:52.2 & -19:00:30 & 15.26 & 1207$\,\pm\,$12 & - & 1297$\,\pm\,$28 & 1207$\,\pm\,$12\\
FS016 & B033604.71-185958.2 & 03:38:19.8 & -18:50:13 & 17.73 & 12691$\,\pm\,$150 & - & 12661$\,\pm\,$17 & background\\
FS020 & NGC 1393 & 03:38:38.6 & -18:25:41 & 13.81 & 2127$\,\pm\,$26 & - & 2117$\,\pm\,$40 & 2127$\,\pm\,$26\\
FS021 & LSBG F548-012 & 03:38:42.1 & -18:53:58 & 17.23 & - & - & 1506$\,\pm\,$66 & 1508$\,\pm\,$66\\
FS023 & NGC 1391 & 03:38:52.9 & -18:21:15 & 14.56$^d$ & 4329$\,\pm\,$39 & 4242$\,\pm\,$46 & - & background\\
FS024 & J03385455-1752565 & 03:38:54.6 & -17:52:57 & 17.37 & - & - & 12473$\,\pm\,$38 & background\\
FS027 & J03390302-1751405 & 03:39:03.0 & -17:51:41 & 15.56 & 12381$\,\pm\,$69 & 12525$\,\pm\,$46 & 12475$\,\pm\,$47 & background\\
FS028 & LSBG F548-011 & 03:39:04.5 & -18:21:37 & 19.22 & - & - & 4442$\,\pm\,$35 & background\\
FS029 & B033649.11-184138.2 $^a$ & 03:39:04.5 & -18:31:56 & 19.16 & - & - & 1978$\,\pm\,$86 & 1978$\,\pm\,$86\\
FS030 & NGC 1394 & 03:39:06.9 & -18:17:32 & 14.69 & 4243$\,\pm\,$16 & - & 4405$\,\pm\,$50 & background\\
FS035 & J03391927-1800359 & 03:39:19.3 & -18:00:36 & 16.46 & 12574$\,\pm\,$41 & 12705$\,\pm\,$46 & 12709$\,\pm\,$32 & background\\
FS036 & B033707.98-185511.4 & 03:39:23.1 & -18:45:30 & 17.28 & - & - & 1817$\,\pm\,$68 & 1817$\,\pm\,$68\\
FS037 & J03392686-1912482 & 03:39:26.8 & -19:12:48 & 17.27 & 19177$\,\pm\,$32 & - & 19188$\,\pm\,$35 & background\\
FS038 & NGC 1402 & 03:39:30.6 & -18:31:37 & 15.17$^d$ & 4184$\,\pm\,$36 & 4356$\,\pm\,$46 & 4297$\,\pm\,$14 & background\\
FS039 & NGC 1400 & 03:39:30.8 & -18:41:17 & 12.30$^e$ & 558$\,\pm\,$14 & 530$\,\pm\,$46 & 617$\,\pm\,$38 & interloper$^b$\\
FS040 & LSBG F548-010 & 03:39:41.6 & -17:55:05 & 18.81 & - & - & 1611$\,\pm\,$98 & 1611$\,\pm\,$98\\
FS051 & LSBG F548-008 & 03:40:04.6 & -17:52:42 & 18.45 & - & - & 9691$\,\pm\,$27 & background\\
FS052 & IC 0343 & 03:40:07.1 & -18:26:36 & 14.64$^d$ & 1841$\,\pm\,$24 & - & 1895$\,\pm\,$36 & 1841$\,\pm\,$24\\
FS053 & LSBG F548-007 & 03:40:10.2 & -18:56:40 & 18.49 & - & - & 1993$\,\pm\,$69 & 1993$\,\pm\,$69\\
FS055 & NGC 1407 & 03:40:11.9 & -18:34:49 & 11.00 & 1779$\,\pm\,$9 & - & 1838$\,\pm\,$58 & 1779$\,\pm\,$9\\
FS054 & J03401339-1819084 & 03:40:13.4 & -18:19:09 & 17.29 & - & - & 1914$\,\pm\,$45 & 1914$\,\pm\,$45\\
FS056 & J03401592-1904544 & 03:40:15.9 & -19:04:54 & 15.83 & 1575$\,\pm\,$61 & 1613$\,\pm\,$46 & 1674$\,\pm\,$38 & 1674$\,\pm\,$38\\
FS057 & ESO 548-G 068 & 03:40:19.2 & -18:55:53 & 14.42$^d$ & 1693$\,\pm\,$18 & - & 1753$\,\pm\,$37 & 1693$\,\pm\,$18\\
FS061 & LSBG F548-006 & 03:40:33.7 & -18:39:03 & 17.70 & - & - & 849$\,\pm\,$75 & 849$\,\pm\,$75\\
FS064 & B033827.69-190614.5 & 03:40:42.6 & -18:56:38 & 19.07 & - & - & 1418$\,\pm\,$84 & 1418$\,\pm\,$84\\
FS065 & J03404323-1838431 & 03:40:43.2 & -18:38:43 & 15.77 & 1308$\,\pm\,$41 & 1373$\,\pm\,$46 & 1394$\,\pm\,$37 & 1394$\,\pm\,$37\\
FS066 & LEDA 074889 & 03:40:46.7 & -18:36:22 & 18.79 & - & - & 31102$\,\pm\,$24 & background\\
FS069 & LSBG F549-038 & 03:40:49.7 & -18:50:50 & 18.21 & - & - & 1266$\,\pm\,$61 & 1266$\,\pm\,$61\\
FS071 & APMBGC 548-110-078 & 03:40:52.5 & -18:28:39 & 16.30 & 1595$\,\pm\,$68 & 1679$\,\pm\,$46 & 1667$\,\pm\,$55 & 1679$\,\pm\,$46\\
FS074 & B033847.86-182654.2 & 03:41:03.4 & -18:17:19 & 18.07 & - & - & 9696$\,\pm\,$28 & background\\
FS075 & ESO 548-G 073 & 03:41:04.4 & -19:05:40 & 15.82 & 989$\,\pm\,$46 & 989$\,\pm\,$46 & 976$\,\pm\,$37 & 976$\,\pm\,$37\\
FS077 & J03410810-1748480 & 03:41:08.1 & -17:48:48 & 17.18 & 19732$\,\pm\,$46 & 19732$\,\pm\,$46 & 19686$\,\pm\,$44 & background\\
FS078 & IC 0345 & 03:41:09.1 & -18:18:51 & 15.34 & 1335$\,\pm\,$33 & 1244$\,\pm\,$46 & 1280$\,\pm\,$36 & 1335$\,\pm\,$33\\
FS081 & J0341186-180206 & 03:41:18.6 & -18:02:06 & 17.11 & - & - & 1505$\,\pm\,$40 & 1505$\,\pm\,$40\\
FS082 & J0341237-183808 & 03:41:23.7 & -18:38:08 & 17.49 & - & - & 1912$\,\pm\,$72 & 1912$\,\pm\,$72\\
FS084 & APMBGC 548-118-089 & 03:41:29.8 & -18:15:50 & 16.44 & - & - & 1870$\,\pm\,$56 & 1870$\,\pm\,$56\\
FS085 & IC 0346 & 03:41:44.6 & -18:16:01 & 14.46 & 2013$\,\pm\,$43 & - & 2085$\,\pm\,$31 & 2085$\,\pm\,$31\\
FS086 & LSBG F549-036 & 03:41:46.6 & -17:44:21 & 17.65 & - & - & 1802$\,\pm\,$85 & 1802$\,\pm\,$85\\
FS088 & ESO 548-G 079 & 03:41:56.1 & -18:53:43 & 15.02 & 2016$\,\pm\,$26 & 2029$\,\pm\,$46 & 2080$\,\pm\,$39 & 2016$\,\pm\,$26\\
FS089 & J03415989-1842469 & 03:41:59.9 & -18:42:47 & 16.30 & - & - & 1846$\,\pm\,$43 & 1846$\,\pm\,$43\\
FS090 & J03420335-1731157 & 03:42:03.3 & -17:31:16 & 15.51 & 11582$\,\pm\,$25 & - & - & background\\
FS092 & LSBG F548-005 & 03:42:13.2 & -18:56:52 & 19.29 & - & - & 1808$\,\pm\,$92 & 1808$\,\pm\,$92\\
FS093 & J03422001-1752117 & 03:42:20.0 & -17:52:12 & 17.01 & - & - & 7133$\,\pm\,$19 & background\\
FS099 & ESO 549-G 002 & 03:42:57.3 & -19:01:12 & 14.90 & 1111$\,\pm\,$10 & - & 1121$\,\pm\,$19 & 1111$\,\pm\,$10\\
FS105 & J03434035-1827314 & 03:43:40.4 & -18:27:32 & 18.54 & - & - & 41312$\,\pm\,$33 & background\\
FS111 & NGC 1440 & 03:45:02.9 & -18:15:58 & 12.15 & 1597$\,\pm\,$27 & - & - & 1597$\,\pm\,$27\\
FS114 & NGC 1452 & 03:45:22.3 & -18:38:01 & 12.48 & 1737$\,\pm\,$10 & - & - & 1737$\,\pm\,$10\\
\\
  & \multicolumn{7}{l}{plus 69 other Ferguson \& Sandage catalogue members with no redshift information}\\
\hline
\end{tabular*}
\begin{flushleft}
\emph{a.} This target was selected as a point source (UCD candidate) rather than as a galaxy.\\
\emph{b.} NGC 1400 has a high velocity relative to the group, although secondary distance measures and a high dark matter content \cite{Quintana..1994} locate it within the group. We treat NGC 1400 as an interloper for dynamical calculations.\\
\emph{c.} NED has a range of velocity measurements between 1948 $\mbox{km} \, \mbox{s}^{-1}$ and 2017 $\mbox{km} \, \mbox{s}^{-1}$ with similar uncertainties. The 6dF DR2 result is intermediate between NED and our 2dF result.\\
\emph{d.} $b_J$ magnitude based on APM galaxy survey data \citep{Maddox..1990}.\\
\emph{e.} Estimated $b_J$ magnitude based on NED data.\\
\end{flushleft}
\end{table*}

\begin{table*}
\caption{NGC 4038 Group (LGG 263)}
\label{table:NGC4038}
\begin{tabular*}{1.00\textwidth}%
     {@{\extracolsep{\fill}}llccccccc} 
\hline
Ref. & Name & RA(J2000) & Dec(J2000) & $b_{j}$ & \multicolumn{3}{|c|}{Velocity ($\mbox{km} \, \mbox{s}^{-1})$} & Revised\\
\cline{6-8}
 &  & [h:m:s] & [d:m:s] &  & NED & 6dF DR2 & 2dF Data & Status\\
(1) & (2) & (3) & (4) & (5) & (6) & (7) & (8) & (9)\\ 
\hline
PGC37325 & NGC 3956 & 11:54:00.7 & -20:34:02 & 12.47 & 1645$\,\pm\,$5 & - & - & 1645$\,\pm\,$5\\
PGC37326 & NGC 3957 & 11:54:01.5 & -19:34:08 & 12.45 & 1637$\,\pm\,$19 & - & - & 1637$\,\pm\,$19\\
PGC37496 & NGC 3981 & 11:56:07.4 & -19:53:46 & 11.89 & 1723$\,\pm\,$4 & - & - & 1723$\,\pm\,$4\\
PGC37565 & ESO 572-23 & 11:56:57.9 & -19:51:18 & 13.39$^a$ & 1805$\,\pm\,$19 & 1771$\,\pm\,$46 & - & 1805$\,\pm\,$19\\
PGC37476 & ESO 572-18 & 11:55:50.6 & -18:11:47 & 13.11 & 1599$\,\pm\,$10 & - & - & 1599$\,\pm\,$10\\
PGC37690 & NGC 4024 & 11:58:31.2 & -18:20:49 & 12.59$^a$ & 1694$\,\pm\,$15 & - & - & 1694$\,\pm\,$15\\
PGC37773 & NGC 4027 & 11:59:30.2 & -19:15:55 & 11.05 & 1671$\,\pm\,$6 & - & 1744$\,\pm\,$10 & 1744$\,\pm\,$10$^c$\\
PGC37967 & NGC 4038 & 12:01:53.0 & -18:52:10 & 10.05 & 1642$\,\pm\,$12 & 1687$\,\pm\,$46 & - & 1642$\,\pm\,$12\\
PGC37969 & NGC 4039 & 12:01:53.6 & -18:53:11 & 10.05 & 1641$\,\pm\,$9 & - & - & 1641$\,\pm\,$9\\
PGC38087 & ESO 572-49 & 12:03:24.4 & -19:31:21 & 14.09 & 1642$\,\pm\,$10 & - & - & 1642$\,\pm\,$10\\
PGC37863 & NGC 4033 & 12:00:34.7 & -17:50:33 & 12.77 & 1617$\,\pm\,$20 & - & - & 1617$\,\pm\,$20\\
PGC37373 & MCG 3-30-19 & 11:54:49.5 & -16:51:50 & 14.60 & 1812$\,\pm\,$6 & - & - & 1812$\,\pm\,$6\\
PGC37320 & NGC 3955 & 11:53:57.1 & -23:09:51 & 12.46 & 1491$\,\pm\,$9 & - & - & 1491$\,\pm\,$9\\
PGC37853 & NGC 4035 & 12:00:29.3 & -15:56:53 & 13.98$^a$ & 1567$\,\pm\,$11 & - & - & 1567$\,\pm\,$11\\
PGC38049 & NGC 4050 & 12:02:53.9 & -16:22:25 & 13.22 & 1761$\,\pm\,$8 & - & - & 1761$\,\pm\,$8\\
PGC38952 & ESO 573-3 & 12:12:55.2 & -20:25:16 & 14.90 & 1546$\,\pm\,$1 & - & - & 1546$\,\pm\,$1\\
PGC37681 & ESO 572-30 & 11:58:25.4 & -22:26:24 & 14.16$^b$ & 1795$\,\pm\,$5 & - & - & 1795$\,\pm\,$5\\
PGC38367 & ESO 505-13 & 12:06:07.2 & -22:50:58 & 14.98 & 1723$\,\pm\,$2 & - & - & 1723$\,\pm\,$2\\
PGC38652 & ESO 505-23 & 12:09:36.2 & -23:24:43 & 15.89 & 1669$\,\pm\,$5 & - & - & 1669$\,\pm\,$5\\
PGC37727 & ESO 572-34 & 11:58:58.1 & -19:01:48 & 13.96 & 1114$\,\pm\,$2 & 1145$\,\pm\,$46 & - & foreground\\
PGC37245 & ESO 572-8 & 11:53:05.5 & -18:22:41 & 14.31 & 1766$\,\pm\,$10 & - & - & 1766$\,\pm\,$10\\
PGC37270 & ESO 572-9 & 11:53:23.0 & -18:10:00 & 16.95$^b$ & 1745$\,\pm\,$3 & - & - & 1745$\,\pm\,$3\\
PGC37513 & ESO 572-22 & 11:56:22.5 & -19:33:07 & 14.40 & 1918$\,\pm\,$10 & - & 1967$\,\pm\,$29 & 1918$\,\pm\,$10\\
PGC37602 & ISZ 59 & 11:57:28.0 & -19:37:27 & 15.23 & 1781$\,\pm\,$41 & - & 2135$\,\pm\,$18 & 2135$\,\pm\,$18\\
PGC37671 & ISZ 60 & 11:58:23.8 & -19:31:03 & 14.80 & 1491$\,\pm\,$86 & 1675$\,\pm\,$46 & - & 1675$\,\pm\,$46\\
PGC37707 & ESO 572-32 & 11:58:45.5 & -19:50:45 & 15.79 & 1638$\,\pm\,$2 & - & - & 1638$\,\pm\,$2\\
PGC37772 & NGC 4027A & 11:59:29.4 & -19:19:55 & 14.58$^a$ & 1747$\,\pm\,$73 & - & - & 1747$\,\pm\,$73\\
- & J11570214-1943413 & 11:57:02.1 & -19:43:42 & 15.52 & 1609$\,\pm\,$46 & - & 1708$\,\pm\,$41 & 1708$\,\pm\,$41\\
\\
\hline
\end{tabular*}
\begin{flushleft}
\emph{a.} $b_J$ magnitudes from SuperCOSMOS bright galaxy measurements (M. Read, private communication).\\
\emph{b.} Estimated $b_J$ magnitude based on NED data.\\
\emph{c.} NED velocities are out of date (reprocessing of old data). 2dF R-values are high.\\
\end{flushleft}
\end{table*}

\begin{table*}
\caption{NGC 4697 Group (LGG 314)}
\label{table:NGC4697}
\begin{tabular*}{1.00\textwidth}%
     {@{\extracolsep{\fill}}llccccccc} 
\hline
Ref. & Name & RA(J2000) & Dec(J2000) & $b_{j}$ & \multicolumn{3}{|c|}{Velocity ($\mbox{km} \, \mbox{s}^{-1})$} & Revised\\
\cline{6-8}
 &  & [h:m:s] & [d:m:s] &  & NED & 6dF DR2 & 2dF Data & Status\\
(1) & (2) & (3) & (4) & (5) & (6) & (7) & (8) & (9)\\ 
\hline
PGC42868 & MCG-1-33-1 & 12:44:03.5 & -05:40:34 & 13.089 & 1430$\,\pm\,$2 & - & - & 1430$\,\pm\,$2\\
PGC43020 & MCG-1-33-3 & 12:45:41.4 & -06:04:08 & 15.72 & 1475$\,\pm\,$4 & - & - & 1475$\,\pm\,$4\\
PGC43276 & NGC 4697 & 12:48:35.9 & -05:48:03 & 11.09 & 1241$\,\pm\,$2 & - & - & 1241$\,\pm\,$2\\
PGC43283 & MCG-1-33-11 & 12:48:43.1 & -05:15:14 & 16.27 & 1341$\,\pm\,$7 & - & - & 1341$\,\pm\,$7\\
PGC43507 & NGC 4731 & 12:51:01.1 & -06:23:35 & 12.31$^c$ & 1489$\,\pm\,$3 & - & - & 1489$\,\pm\,$3\\
PGC43697 & MCG-1-33-33 & 12:52:36.3 & -06:17:20 & 14.36$^c$ & 1534$\,\pm\,$5 & - & - & 1534$\,\pm\,$5\\
PGC43786 & UGCA 305 & 12:53:21.4 & -04:58:41 & 16.59 & 1411$\,\pm\,$6 & - & - & 1411$\,\pm\,$6\\
PGC43826 & NGC 4775 & 12:53:45.7 & -06:37:20 & 12.60$^d$ & 1568$\,\pm\,$3 & - & - & 1568$\,\pm\,$3\\
PGC44278 & MCG-1-33-59 & 12:57:16.5 & -05:20:45 & 14.78 & 1258$\,\pm\,$3 & - & - & 1258$\,\pm\,$3\\
PGC44264 & UGCA 310 & 12:57:12.1 & -04:09:32 & 15.40 & 1529$\,\pm\,$4 & - & - & 1529$\,\pm\,$4\\
PGC44166 & IC 3908 & 12:56:40.4 & -07:33:40 & 13.66 & 1296$\,\pm\,$4 & - & - & 1296$\,\pm\,$4\\
PGC44506 & MCG-1-33-61 & 12:58:48.9 & -06:06:46 & 14.86 & 1600$\,\pm\,$3 & - & - & 1600$\,\pm\,$3\\
PGC45165 & NGC 4941 & 13:04:13.1 & -05:33:06 & 12.79 & 1108$\,\pm\,$5 & - & - & 1108$\,\pm\,$5\\
PGC45246 & NGC 4951 & 13:05:07.7 & -06:29:38 & 12.68 & 1176$\,\pm\,$4 & - & - & 1176$\,\pm\,$4\\
PGC45224 & NGC 4948 & 13:04:55.9 & -07:56:52 & 13.67 & 1330$\,\pm\,$10 & - & - & 1330$\,\pm\,$10\\
PGC45254 & MCG-1-33-82 & 13:05:14.3 & -07:53:21 & 16.62 & 1123$\,\pm\,$11 & - & - & 1123$\,\pm\,$11\\
PGC45313 & NGC 4958 & 13:05:48.9 & -08:01:13 & 12.10 & 1455$\,\pm\,$9 & - & - & 1455$\,\pm\,$9\\
PGC43341 & MCG-1-33-14 & 12:49:18.3 & -04:00:59 & 16.24 & 1506$\,\pm\,$4 & - & - & 1506$\,\pm\,$4\\
- & LCRS B125056.4-053926 $^a$ & 12:53:31.6 & -05:55:43 & 16.87 & 1071$\,\pm\,$4 & - & 1093$\,\pm\,$18 & 1071$\,\pm\,$4\\
- & LCRS B124921.9-062421 & 12:51:57.3 & -06:40:39 & 16.26 & - & - & 1450$\,\pm\,$8 & 1450$\,\pm\,$8\\
- & B125228.28-060555.8 $^b$ & 12:55:03.7 & -06:22:10 & 15.47 & - & - & 1329$\,\pm\,$24 & 1329$\,\pm\,$24\\
- & NGC 4731A & 12:51:13.3 & -06:33:34 & 14.64 & 1497$\,\pm\,$37 & - & 1514$\,\pm\,$12 & 1514$\,\pm\,$12\\
- & NGC4813 & 12:56:36.1 & -06:49:05 & 14.16 & 1373$\,\pm\,$39 & - & 1394$\,\pm\,$37 & 1394$\,\pm\,$37\\
- & LCRS B125056.4-053926 & 12:53:31.4 & -05:55:33 & 16.87 & 1071$\,\pm\,$4 & - & 1118$\,\pm\,$22 & 1071$\,\pm\,$4\\
\\
\hline
\end{tabular*}
\begin{flushleft}
\emph{a.} LCRS refers to the Las Campanas Redshift Survey (see http://qold.astro.utoronto.ca/~lin/lcrs.html).\\
\emph{b.} This is a galaxy pair.\\
\emph{c.} $b_J$ magnitude based on APM galaxy survey data \citep{Maddox..1990}.\\
\emph{d.} $b_J$ magnitude from SuperCOSMOS bright galaxy measurements (M. Read, private communication).\\
\end{flushleft}
\end{table*}

\begin{table*}
\caption{NGC 5084 Group (LGG 345)}
\label{table:NGC5084}
\begin{tabular*}{1.00\textwidth}%
     {@{\extracolsep{\fill}}llccccccc} 
\hline
Ref. & Name & RA(J2000) & Dec(J2000) & $b_{j}$ & \multicolumn{3}{|c|}{Velocity ($\mbox{km} \, \mbox{s}^{-1})$} & Revised\\
\cline{6-8}
 &  & [h:m:s] & [d:m:s] &  & NED & 6dF DR2 & 2dF Data & Status\\
(1) & (2) & (3) & (4) & (5) & (6) & (7) & (8) & (9)\\ 
\hline
PGC46525 & NGC 5084 & 13:20:16.9 & -21:49:39 & 11.05 & 1721$\,\pm\,$3 & 1660$\,\pm\,$46 & - & 1721$\,\pm\,$3\\
PGC46541 & NGC 5087 & 13:20:24.9 & -20:36:40 & 12.12 & 1832$\,\pm\,$56 & 1819$\,\pm\,$46 & 1788$\,\pm\,$87 & 1819$\,\pm\,$46\\
PGC46938 & NGC 5134 & 13:25:18.5 & -21:08:03 & 11.68 & 1757$\,\pm\,$6 & 1792$\,\pm\,$46 & 1677$\,\pm\,$44 & 1757$\,\pm\,$6\\
PGC46889 & ESO 576-50 & 13:24:42.1 & -19:41:50 & 13.07 & 1984$\,\pm\,$7 & - & - & 1984$\,\pm\,$7\\
PGC46574 & ESO 576-40 & 13:20:43.7 & -22:03:04 & 14.53 & 2080$\,\pm\,$5 & 2096$\,\pm\,$46 & - & 2080$\,\pm\,$5\\
- & - & 13:21:36.2 & -20:36:35 & 16.71 & - & - & 2184$\,\pm\,$16 & 2184$\,\pm\,$16\\
- & - & 13:23:30.4 & -20:28:44 & 16.69 & - & - & 1574$\,\pm\,$18 & 1574$\,\pm\,$18\\
- & - & 13:22:15.7 & -21:50:31 & 16.64 & - & - & 1863$\,\pm\,$11 & 1863$\,\pm\,$11\\
- & J1317585-211025 & 13:17:58.3 & -21:10:28 & 16.48 & - & - & 1575$\,\pm\,$21 & 1575$\,\pm\,$21\\
- & ESO 576- G 031 & 13:19:47.2 & -21:54:00 & 13.84 & 1432$\,\pm\,$28 & - & 1389$\,\pm\,$42 & 1432$\,\pm\,$28\\
- & IC 4237 & 13:24:32.7 & -21:08:14 & 12.56 & 2643$\,\pm\,$5 & - & 2669$\,\pm\,$175 & background $^a$\\
- & NGC 5068 & 13:18:54.6 & -21:02:20 & 10.71 & 668$\,\pm\,$3 & - & 649$\,\pm\,$20 & foreground $^a$\\
- & [CCF97] G4 & 13:20:12.6 & -22:00:11 & 18.40 & 1092$\,\pm\,$123 & - & 1690$\,\pm\,$20 & 1690$\,\pm\,$20\\
- & - & 13:22:39.5 & -21:48:33 & 17.44 & - & - & 2067$\,\pm\,$19 & 2067$\,\pm\,$19\\
- & [CCF97] G8 & 13:20:49.6 & -22:03:19 & 16.96 & 2080$\,\pm\,$6 & - & 2108$\,\pm\,$12 & 2080$\,\pm\,$6\\
\\
\hline
\end{tabular*}
\begin{flushleft}
\emph{a.} These bright galaxies are isolated from the relatively compact velocity range of the other group galaxies.\\
\end{flushleft}
\end{table*}

\label{lastpage}

\end{document}